\documentclass[10pt,journal,compsoc]{IEEEtran}
\ifCLASSOPTIONcompsoc
  \usepackage[nocompress]{cite}
\else
  \usepackage{cite}
\fi
\ifCLASSINFOpdf
\else
\fi
\usepackage{cite}
\usepackage{amsmath}
\usepackage{amssymb}
\usepackage{graphicx}

\usepackage{threeparttable}
\usepackage{multirow}
\usepackage{booktabs}

\usepackage{array}

\usepackage{xcolor}
\usepackage{colortbl}

\usepackage{subfigure}
\usepackage{caption}

\usepackage{rotating}
\usepackage{url}

\begin{document}

\definecolor{light_grey}{rgb}{.90234375,.8984375,.8984375}

\newcommand{\tabincell}[2]{\begin{tabular}{@{}#1@{}}#2\end{tabular}}
\newcommand{\PreserveBackslash}[1]{\let\temp=\\#1\let\\=\temp}
\newcolumntype{C}[1]{>{\PreserveBackslash\centering}p{#1}}
\newcolumntype{R}[1]{>{\PreserveBackslash\raggedleft}p{#1}}
\newcolumntype{L}[1]{>{\PreserveBackslash\raggedright}p{#1}}
\newcommand{\eight}{\fontsize{8}{\baselineskip}\selectfont}

%
\title{Bridging Semantic Gaps between Natural Languages and APIs with Word Embedding}
%
%
%
%

\author{Xiaochen~Li,
        He~Jiang,~\IEEEmembership{Member,~IEEE,}
        Yasutaka~Kamei,~\IEEEmembership{Member,~IEEE,}
        and~Xin~Chen,
\IEEEcompsocitemizethanks{\IEEEcompsocthanksitem X. Li and X. Chen are with School of Software, Dalian University of Technology, Dalian, China. E-mail: li1989@mail.dlut.edu.cn, chenxin4391@mail.dlut.edu.cn
\IEEEcompsocthanksitem H. Jiang is with School of Software, Dalian University of Technology, Dalian, China, and Key Laboratory for Ubiquitous Network and Service Software of Liaoning Province. E-mail: jianghe@dlut.edu.cn
\IEEEcompsocthanksitem Y. Kamei is with the Principles of Software Languages Group (POSL), Kyushu University, Japan. Email: kamei@ait.kyushu-u.ac.jp}}

%
%

\markboth{}%
{Shell \MakeLowercase{\textit{et al.}}: Bare Demo of IEEEtran.cls for IEEE Journals}
%



\IEEEtitleabstractindextext{%
\begin{abstract}

Developers increasingly rely on text matching tools to analyze the relation between natural language words and APIs.
However, semantic gaps, namely textual mismatches between words and APIs, negatively affect these tools.
Previous studies have transformed words or APIs into low-dimensional vectors for matching;
however, inaccurate results were obtained due to the failure of modeling words and APIs simultaneously.
To resolve this problem, two main challenges are to be addressed:
the acquisition of massive words and APIs for mining and the alignment of words and APIs for modeling.
Therefore, this study proposes Word2API to effectively estimate relatedness of words and APIs.
Word2API collects millions of commonly used words and APIs from code repositories to address the acquisition challenge.
Then, a shuffling strategy is used to transform related words and APIs into tuples to address the alignment challenge.
Using these tuples, Word2API models words and APIs simultaneously.
Word2API outperforms baselines by 10\%-49.6\% of relatedness estimation in terms of precision and NDCG.
Word2API is also effective on solving typical software tasks, e.g., query expansion and API documents linking.
A simple system with Word2API-expanded queries recommends up to 21.4\% more related APIs for developers.
Meanwhile, Word2API improves comparison algorithms by 7.9\%-17.4\% in linking questions in Question\&Answer communities to API documents.
\end{abstract}

}

\maketitle

\IEEEdisplaynontitleabstractindextext

%
\IEEEpeerreviewmaketitle

\section{Introduction}
%
%
%
%


\IEEEPARstart{S}oftware developers put considerable efforts to study APIs (Application Programming Interfaces) \cite{nie2016query,Robillard2011A}.
To facilitate this process, many tools have been developed to retrieve information about APIs,
e.g., searching API sequences based on a query \cite{raghothaman2016swim} or recommending API documents for answering technical questions \cite{ye2016word}.
These tools generally utilize information retrieval models, such as Vector Space Model (VSM) \cite{Baeza2011Modern},
to transform queries and APIs into words
and conduct text matching to find required APIs or API documents \cite{lv2015codehow}.
Since there is usually a mismatch between the content of natural languages and APIs,
the performance of these tools is negatively affected \cite{lv2015codehow}.

For example, in the task of API sequences recommendation,
when a developer searches for APIs implementing `generate md5 hash code',
Java APIs of `MessageDigest\#getInstance' and `MessageDigest\#digest' may be required \cite{gu2016deep}.
However, neither the word `md5' nor `hash code' could be matched with these APIs,
which misleads information retrieval models to return the required APIs \cite{lv2015codehow}.

Another example is from
the task of API documents linking.
Developers usually ask technical questions on Question \& Answer communities,
e.g., `How to (conduct a) sanity check (on) a date in Java'.\footnote{https://stackoverflow.com/questions/226910/}
In their answers, the API `Calendar\#setLenient' is recommended by participants.
However, based on text matching,
the relationship between `sanity check (on) a date' and `Calendar\#setLenient' is difficult to be determined.
The question submitter even complained that `(it is) not so obvious to use lenient calendar'.

In the above examples,
the mismatches between natural language words and APIs are semantic gaps.
The gaps hinder developers from using APIs \cite{piccioni2013empirical}
and tend to bring thousands of defects in API documents \cite{zhou2017analyzing}.
They are also a major obstacle for the effectiveness of many software engineering tools \cite{lv2015codehow, mahmoud2015estimating}.
Previous studies have shown that a text-matching based retrieval tool could only return 25.7\% to 38.4\% useful code snippets in top-10 results for developers' queries \cite{lv2015codehow}.
To bridge the gaps, a fundamental solution is to correctly estimate the relatedness or similarity between a word and an API or a set of words and APIs,
e.g., generating accurate similarity between words `sanity check (on) a date' and the API 'Calendar\#setLenient'.

Motivated by the aim of achieving such a solution,
many algorithms for relatedness estimation have been proposed,
including latent semantic analysis \cite{landauer1997solution}, co-occurrence analysis \cite{mahmoud2015estimating}, WordNet thesaurus \cite{miller1995wordnet}, etc.
Among them, word embedding has recently shown its advantages \cite{nguyen2017exploring, ye2016word};
it constructs low-dimensional vectors of words or APIs for relatedness estimation.
Existing studies tried to train software word embedding \cite{ye2016word} based on Java/Eclipse tutorials and user guides, as well as API embedding \cite{nguyen2017exploring} with API sequences from different programming languages.
These strategies may still be ineffective to estimate the words-APIs relatedness,
as they only learn the relationships for either words or APIs.

To improve the performance of existing solutions,
it is necessary to model the words and APIs simultaneously into the same vector space.
However, two main challenges are to be addressed:
the acquisition challenge and the alignment challenge.
The acquisition challenge is how to collect a large number of documents that contain diverse words and APIs.
API tutorials and user guides are usually full of words, but have few APIs.
The alignment challenge is how to align words and APIs to fully mine their overall relationship in a fixed window size,
since
word embedding mines word-API relationships based on the co-occurrence of words and APIs.

In this study we propose Word2API to address the two challenges.
Word2API first collects large-scale files with source code and method comments from GitHub\footnote{GitHub. https://github.com/} to address the acquisition challenge.
Source code and method comments usually contain diverse words and APIs commonly used by developers.
Then, Word2API preprocesses these files.
It extracts words in method comments and APIs in source code with a set of heuristic rules,
which are efficient in identifying semantically related words and APIs in the files.
After that, the extracted words and APIs regarding the same method are combined as a word-API tuple.
Since the method comment always comes before the API calls in a method\footnote{In this paper, `method' refers to a function or procedure defined in a class. We use `algorithm' or `approach' to describe Word2API},
the co-occurrence of words and APIs may be hardly mined in a fixed window.
Word2API leverages a shuffling strategy to address the alignment challenge.
This strategy randomly shuffles words and APIs in a word-API tuple to form a shuffled tuple for training.
Since there is valuable information among all words and APIs in the same word-API tuple,
this strategy is effective in increasing the word-API collocations and revealing the overall relationship between words and APIs in a fixed window.
Finally, Word2API applies word embedding on the shuffled results to generate word and API vectors.

We trained Word2API with 391 thousand Java projects consisting of more than 31 million source code files from GitHub.
Word2API generates vectors for 89,422 words and 37,431 APIs.
We evaluate Word2API by recommending semantically related APIs for a word.
For 31 out of 50 words, the top-1 recommended API is related,
which outperforms comparison algorithms by 10\%-49.6\% in terms of precision and Normalized Discounted Cumulative Gain (NDCG).
Meanwhile, the shuffling strategy significantly improves the effectiveness of word embedding in constructing semantically related vectors from word-API tuples.

Besides, we demonstrate two applications of Word2API, including query expansion for API sequences recommendation \cite{raghothaman2016swim} and API documents linking \cite{ye2016word}.
API sequences recommendation recommends API sequences in source code for a user query.
API documents linking links questions in Q\&A communities to the API documents that may be useful to answer the questions.
For the first task, Word2API expands a user query into a set of APIs.
A simple system with Word2API-expanded queries can recommend up to 21.4\% more related API sequences than baseline algorithms.
For the second task,
Word2API outperforms existing algorithms by 8.9\% and 7.9\% in linking useful API documents to questions in Stack Overflow
in terms of Mean Average Precision (MAP) and Mean Reciprocal Rank (MRR) respectively.

To conclude, we make the following contributions.

\begin{enumerate}
    \item We propose Word2API to solve the problem of constructing low-dimensional representations for both words and APIs simultaneously.
        Word2API successfully addresses the acquisition challenge and alignment challenge in this problem.

    \item With Word2API, we generate 126,853 word and API vectors to bridge the sematic gaps between natural language words and APIs.
    We publish the generated vectors as a dictionary for research.\footnote{The dictionary W2A$_\textrm{DIC}$. https://github.com/softw-lab/word2api}

    \item We show two applications of Word2API.
        Word2API improves the performance of two typical software engineering tasks, i.e., API sequences recommendation and API documents linking.
\end{enumerate}

\textbf{Outline}.
Section \ref{sec:background} presents the background of this study.
Section \ref{sec:word2api_model} shows the framework of Word2API.
Experimental settings and results on relatedness estimation are introduced in Sections \ref{sec:evaluation_setting} and \ref{sec:evalution_results} respectively.
Two applications of Word2API are shown in Sections \ref{sec:app1} and \ref{sec:app2_linking}.
In Section \ref{sec:threats}, threats to validity are discussed.
We review the related work in Section \ref{sec:related_work}.
Finally, Section \ref{sec:conclusion} concludes this paper.

\section{Background}
\label{sec:background}
\subsection{Terminology}
This subsection defines the major terms used in this paper.

\emph{APIs} are pre-defined functions for communication between software components \cite{apidefinition}.
They are designed under the criteria of high readability, reusability, extendibility, etc. \cite{bloch2006how}.
In this study, an API refers to a method-level API that consists of the fully qualified name of an API type and a method name.

A \emph{word} is a natural language element in a document or text to express human intentions.
We take all the non-API elements in a document or text as words.
In software engineering, there are many API-like words \cite{ye2016word} such as `readLine', `IOException', etc.
We also call them words.

In addition, `\emph{term}' is used to generally indicate either \emph{APIs} or natural language \emph{words}.

We use the word `\emph{document}' to indicate a text with many words or APIs.
Some special documents in software engineering are \emph{API documents} \cite{ye2016word}.
In this study, API documents refer to the documents in API specifications.
Each API document contains method-level APIs in the same class and illustrates the class-description, method-description, etc.

\subsection{Word Embedding}
\begin{figure}

\centering
  \includegraphics[width=8.68cm]{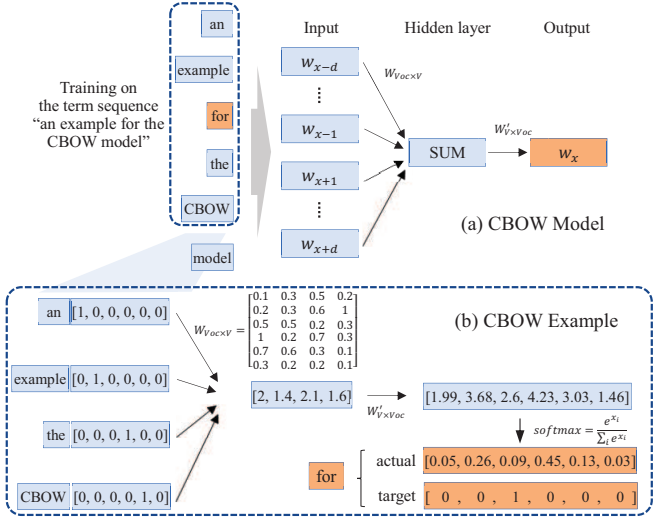}
  \caption{CBOW model for word embedding.}
  \label{fig:cbow}
\end{figure}
\begin{figure*}[htb]
\centering
  \includegraphics[width=17.6cm]{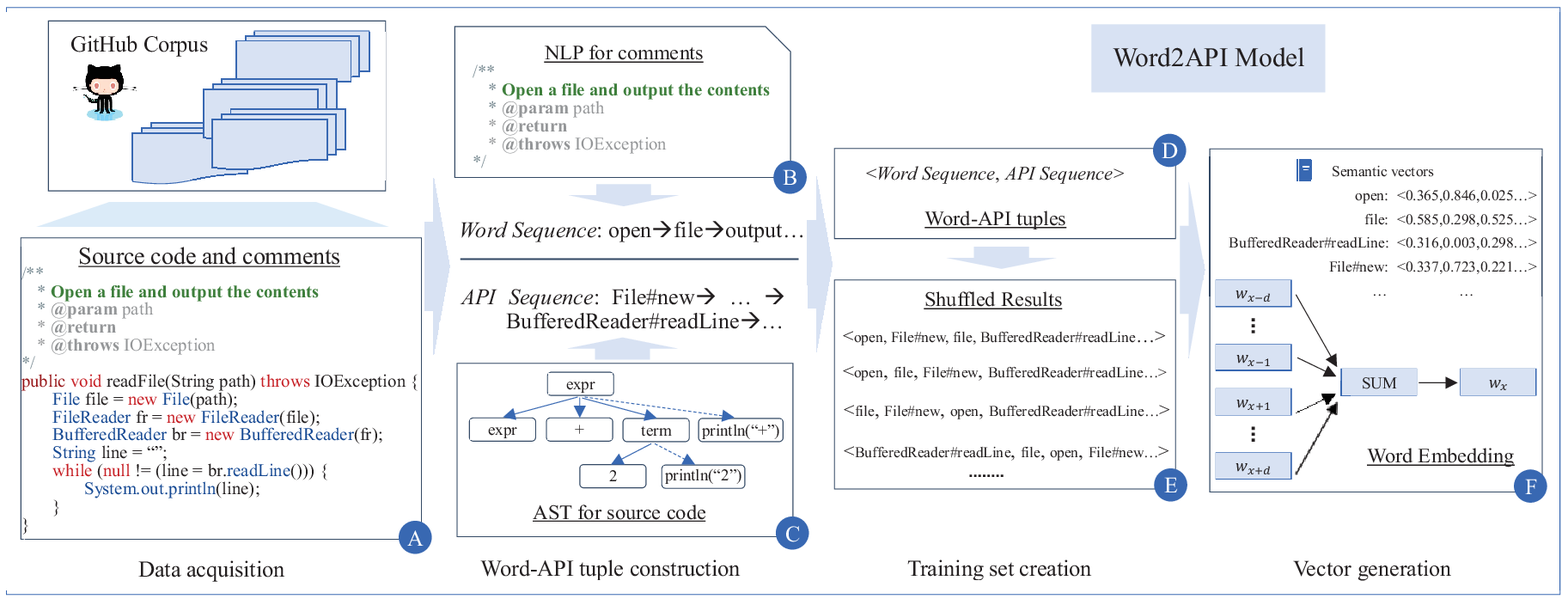}
  \caption{Framework of Word2API model. (A) crawls the source code and comments from GitHub. (B) and (C) extract the word sequences and API sequences in the crawled data.
  (D) combines the sequences as a tuple. (E) shuffles the tuples to generate an unlabeled training set. (F) applies word embedding on the training set to get the term vectors.}
  \label{fig:framework}
\end{figure*}

Word embedding is a fundamental component of Word2API.
It was originally designed to transform words in word sequences into low-dimensional vectors \cite{mikolov2013distributed}.
Many models have been proposed to implement word embedding, e.g., Continuous Bag-of-Words model (CBOW) \cite{mikolov2013efficient}, continuous Skip-gram model (Skip-gram) \cite{mikolov2013distributed}, etc.
To facilitate the use of these models, Google publishes a tool\footnote{Google tool. https://code.google.com/archive/p/word2vec/}
that implements the CBOW and Skip-gram models.
We take the CBOW model as an example to explain word embedding,
as it is the default model in the word embedding tool.

CBOW is a neural network model to learn word representations from an unlabeled training set \cite{mikolov2013efficient}.
Fig. \ref{fig:cbow}(a) presents the framework of CBOW.
CBOW consists of an input layer, an output layer, and a hidden layer.
The hidden layer $h$ is a $1 \times V$ vector to represent words in a low-dimensional space.
$V$ is pre-defined by users.
CBOW uses a matric $W_{Voc\!\times\! V}$ to propagate information between layers,
where $Voc$ is the vocabulary of the training set.

Initially, we randomly initialize the values of $W_{Voc\!\times\! V}$ and represent each word $x$ in $Voc$ with a one-hot vector $w_x$.
The one-hot vector is a zero vector with the exception of a single 1 to uniquely identify the word (Fig. \ref{fig:cbow}(b)).
The vector length is the same as the vocabulary size $|Voc|$.

With these one-hot vectors, CBOW tries to predict the center word with its surrounding context in a fixed window size $d$.
Specifically, CBOW takes in the vectors of the surrounding words \begin{small}$ W_x^d = \{w_{x-d},\dots,w_{x-1},w_{x+1},\dots,w_{x+d}\}$\end{small} in a 2$d$ sized window as the input
and the vector of the center word \begin{small}$ w_{x}$\end{small} as the target output.
For example, if $d=2$, $V=4$ and `for' is the center word,
then the input includes the vectors of `an', `example', `the', `CBOW'.
Based on $W_{Voc\!\times\! V}$,
CBOW propagates the input to the hidden layer $h$
\begin{equation}
\begin{small}
\begin{aligned}
h\!\!=\!\!\frac{1}{2d}(w_{x\!-\!d}\!+\!\!\dots\!\!+\!w_{x\!-\!1}\!+\!w_{x\!+\!1}\!\!+\!\dots\!+\!w_{x\!+\!d})\!\cdot\! W_{Voc\!\times\! V}
\end{aligned}
\end{small}
\end{equation}

Then, the vector in $h$ continues forward propagating according to the parameter matrices $W'_{Voc\!\times\! V}$:
\begin{equation}
\begin{small}
\begin{aligned}
w_{1 \times Voc} = softmax(h \cdot W'_{Voc\!\times\! V}),
\end{aligned}
\end{small}
\end{equation}
where $w_{1\times Voc}$ is the actual output of the center word.
For example, the network outputs a vector [0.05, 0.26, 0.09, 0.45, 0.13, 0.03] in Fig. \ref{fig:cbow}(b).
Since $w_{1\times Voc}$ is far different from the target output $w_x$=[0, 0, 1, 0, 0, 0],
CBOW aims to maximize the average probability that the actual output is $w_x$:
\begin{equation}
\begin{small}
\begin{aligned}
L_M=\frac{1}{X}\sum_{x=1}^{X}\log p(w_x | W_x^d)
\end{aligned}
\end{small}
\end{equation}


CBOW optimizes the output by tuning the parameter matrix \begin{small}$W_{Voc\times V}$\end{small} with back propagation.
After training, we get the values of the final parameter matrix.
For a word x, the low-dimensional vector is calculated as $w_x \cdot W_{Voc\!\times\! V}$.

\section{The Word2API Model}
\label{sec:word2api_model}

Word2API represents natural language words and APIs with low-dimensional vectors.
As depicted in Fig. \ref{fig:framework}, Word2API consists of four steps, including data acquisition, word-API tuple construction, training set creation, and vector generation.
We detail these steps in this section.

\subsection{Data Acquisition}
To train the vectors for words and APIs, we construct a large-scale corpus with source code and method comments.
The corpus (referred as GitHub corpus) is constructed from the Java projects created from 2008 to 2016 on GitHub.
We analyze Java projects
as they have a broad impact on software development.
However, Word2API is independent of programming languages.
We download the zip snapshots in July 2017 of these projects with GitHub APIs.\footnote{GitHub APIs. https://developer.github.com/v3/}
We exclude the projects with zero stars, since they are usually toy or experimental projects \cite{gu2016deep}.
For each project, all the Java files are extracted.
Each file consists of several methods and their comments (Fig. \ref{fig:framework}(A)).
In total, we collect 391,690 Java projects with 31,211,030 source code files.
It should be noted that, in GitHub,
a project may have many forks or third-party source code \cite{Kalliamvakou2016An},
leading to duplicate code snippets.
We keep these duplications in the data set, as forking projects is a basic characteristic of GitHub.


\subsection{Word-API Tuple Construction}
\label{label:word-api_tuples}

With the GitHub corpus, we construct word-API tuples.
A word-API tuple is a combination of a set of words and the corresponding APIs.
We construct the tuples by analyzing the source code of these Java projects.

Specifically, we construct an AST (Abstract Syntax Tree) for each method in the source code by Eclipse's JDT Core Component.\footnote{Eclipse JDT Core Component. http://www.eclipse.org/jdt/core/}
In the AST, we extract the method comment (Fig. \ref{fig:framework}(B)) and its corresponding API types and method calls in the method body (Fig. \ref{fig:framework}(C)) to construct a word-API tuple.
The word-API tuple consists of a word sequence extracted from the method comment and an API sequence obtained from API types and method calls in the method body.

For the method comment, we remove the HTML tags that match the regular expression `$<$.*?$>$',
and split the sentences in the method comment by `. '.
In Java language, sentences in a method comment are typically enclosed between `/**' and `*/' above the method body.
We extract the words in the first sentence to make up the word sequence portion of a word-API tuple,
since this sentence is usually a semantically related high-level summary of a method \cite{gu2016deep}.

For the method body, we extract Java Standard Edition (SE) API types and method calls to make up the API sequence portion of the word-API tuple.
We note that a method is usually implemented with many syntactic units \cite{nguyen2017exploring},
including APIs, variables/identifiers, literals, etc.
Java SE APIs may not fully reveal the intents of a method comment.
However, they are still semantically related to the comment \cite{gu2016deep}.
We extract Java SE APIs as follows:

\begin{itemize}
	\item We traverse the AST of a method to collect the APIs for class instance creation and method calls.
	We represent these APIs with their fully qualified names by resolving the method binding.
	If an API is the argument of another API,
    we represent the API in the argument list first.
	For example, `BufferedReader br = new BufferedReader(new FileReader()); br.readLine())'
    is represented as `java.io.FileReader\#new, java.io.BufferedReader\#new, and java.io.BufferedReader\#readLine'.
	We omit the return type and argument types in this representation,
	since the overloaded APIs of different return types or argument types usually convey the same semantic meaning \cite{javaoverriding}.
	\item We extract Java SE APIs from the collected APIs
    by matching their package names with the ones in the
    Java SE API specification\footnote{Java SE API Spec. http://docs.oracle.com/javase/8/docs/api/} (also called API references).
	We delete the tuples without Java SE APIs.
\end{itemize}

After the above process, a set of word-API tuples are achieved.
We assume that the word sequence in each tuple summarizes the behaviors or purposes of the corresponding APIs.
However,
besides summarizing APIs,
developers may also add TODO lists, notes, etc. in the method comments \cite{pascarella2017classifying},
which are noises in our scenario.
Therefore, we filter out these tuples, if the word sequence in a tuple:

\begin{itemize}
    \item starts with `TODO', `FIXME', `HACK', `REVISIT', `DOCUMENTME', `XXX';
    these tags are commonly used for task annotations instead of summarizing APIs \cite{margaret30todo},
    e.g., `TODO remove this';

    \item starts with words like `note', `test';
    developers use these words to write an explanatory or auxiliary comments \cite{gu2016deep, howard2013automatically},
    e.g., `testing purpose only';



    \item is a single word instead of a meaningful sentence.

\end{itemize}

For the remaining word sequences, we perform
tokenization \cite{lotufo2015modelling}, stop words removal\footnote{Default English stop words. http://www.ranks.nl/stopwords} and stemming \cite{porter1980algorithm}.
We remove words that are numbers or single letters.
If a word is an API-like word, we split it according to its camel style, e.g., splitting `nextInt' into `next' and `int'.
Finally, 13,883,230 tuples are constructed (Fig. \ref{fig:framework}(D)).

\subsection{Training Set Creation}
\label{sec:training_set_creation}

This step creates an unlabeled training set with the constructed word-API tuples for word embedding.
Word embedding is a co-occurrence based method that analyzes the relationship of terms in a fixed window size.
Word embedding works well in a monolingual scenario, e.g., sequential natural language words \cite{ye2016word}, source code identifiers \cite{allamanis2015suggesting}, and API sequences \cite{nguyen2017exploring},
since words or APIs nearby have strong semantic relatedness.
In contrast, it may be hard for word embedding to capture the co-occurrences between words and APIs in a bilingual scenario
such as comments and their corresponding APIs.
In this scenario, words and APIs usually do not appear within each other's window,
e.g., words in the method comments always come before the APIs.
The problem mainly comes from the word-API tuples we collected.
However, to the best of our knowledge,
no training set could be directly used for effectively mining word embedding for both words and APIs like in a monolingual scenario.
An ideal training set should both have a large number of words and APIs and properly align semantic relatedness collocations of words and APIs.
Since word-API tuples consist of diverse words and APIs frequently used by developers,
the remaining challenge is, how to align words and APIs into a fixed window for relationship mining.

To resolve this problem, we merge words and APIs in the same tuple together
and randomly shuffle them to create the training set.
The shuffling step is to break the fixed location of words and APIs.
It tries to obtain enough collocations between each word/API and other APIs/words.
To increase semantically related collocations,
we repeat the shuffling step ten times to generate ten shuffled copies of an original word-API tuple.
Fig. \ref{fig:framework}(E) is the shuffled results of the word-API tuple created from Fig. \ref{fig:framework}(B) and Fig. \ref{fig:framework}(C).
After shuffling, words and APIs tend to co-occur in a small window.
We take these shuffled results as the training set for word embedding.
The training set contains 138,832,300 shuffled results.
Its size is more than 30 gigabyte.

The underlying reason of the above procedure is that
words and APIs in the same word-API tuple tend to contain valuable semantic information (relatedness) for mining.
The shuffling strategy increases the information interaction and helps word embedding learn the knowledge of collocations between words and APIs in a tuple.
After shuffling, the collocations of words and APIs increase,
i.e., words and APIs have higher chances to appear within each other's window.
Hence, word embedding could learn the overall knowledge of each tuple.
Since the shuffling is random,
we repeat the shuffling step to increase related word-API collocations.
We evaluate the shuffling step in Section \ref{sec:rq3}.


\subsection{Vector Generation}
\label{label:vectors_generation}

The last step of Word2API is to train a word embedding model with the training set for vector generation.
We utilize the word embedding tool for unsupervised training.
Word embedding models have many parameters, e.g., `window size', `vector dimension', etc.
Although previous studies show that task-specific parameter optimization influences algorithm performance \cite{tantithamthavorn2016automated},
such optimization may threaten the generalization of an algorithm.
Hence, in this study, all the parameters in the tool are set to the default ones
except the `-min-count' (the threshold to discard a word or API).
Since we generate ten shuffled results for a tuple,
the parameter `-min-count' is set to 50 instead of the default value of 5.
It means that we discard all the words and APIs that appear less than 50 times in the training set.
For some important parameters,
we train word embedding with the default model CBOW,
a more efficient model compared to the Skip-gram model in the word embedding tool\footnote{We compare CBOW and Skip-gram in Sec. S1 of the supplement. }.
The default window size is 5 and the dimension of the generated vectors is 100.
The window size determines how many words or APIs nearby are considered as co-occurred
and the vector dimension reflects the dimension of the generated vector for each word or API.
The other parameters are listed as follows:

- `sample' is 1e-3: the threshold to down-sample a high-frequency term.
The word embedding tool down-samples a term $t_i$ in the training set by $P(t_i)=(\sqrt{ z(t_i) / sample}+1)\times sample / z(t_i)$\footnote{\url{https://github.com/dav/word2vec/blob/master/src/word2vec.c}}, where $z(t_i)$ is the probability that term $i$ appears in the training set and $P(t_i)$ is the probability to keep this term in the training set. When a term appears frequently, $P(t_i)$ tends to be small, which means the probability to keep this term in the training set is low.

- `hs' is 0: hierarchical softmax is not used for training.

- `negative' is 5: the number of random-selected negative samples in a window.

- `iter' is 5: the number of times to iterate the training set.

- `alpha' is 0.05: the starting learning rate.

- `thread' is 32: the number of threads for training.

After running the word embedding tool, 89,422 word vectors and 37,431 API vectors are generated eventually.
These vectors are important to bridge the semantic gaps between natural language words and APIs.
For this purpose, we define word-API similarity and words-APIs similarity:

\emph{Word-API Similarity}
is the similarity between a word $w$ and an API $a$. It is the cosine similarity of vectors \begin{small}$\overrightarrow{V_w}$\end{small} and \begin{small}$\overrightarrow{V_a}$\end{small}:
\begin{equation}
\label{fun:word_api_similarity}
\begin{small}
\begin{aligned}
sim(w, a) = \frac{\overrightarrow{V_w}\cdot\overrightarrow{V_a}}{|\overrightarrow{V_w}||\overrightarrow{V_a}|}.
\end{aligned}
\end{small}
\end{equation}

\emph{Words-APIs Similarity}
extends \emph{Word-API Similarity} to a set of words $W$ and a set of APIs $A$ \cite{mihalcea2006corpus}:
\begin{equation}
\label{fun:words_apis_similarity}
\begin{scriptsize}
\begin{aligned}
sim(W\!,\! A)\!\! =\!\! \frac{1}{2}\!\!\left(\!\!\frac{\sum\!\!{(sim_{max}(w, A)\!\!\times\!\! idf(w))}}{\sum{idf(w)}}
\!\!+\!\! \frac{\sum\!\!{(sim_{max}(a, W)\!\!\times\!\! idf(a))}}{\sum{idf(a)}} \!\!\right),
\end{aligned}
\end{scriptsize}
\end{equation}
where $sim_{max}(w, A)$ returns the highest similarity between $w$ and each API $a\in A$,
and $idf(w)$ is calculated as the number of documents (word sequences in word-API tuples) divided by the number of documents that contain $w$.
Similarly, $sim_{max}(a, W)$ and $idf(a)$ can be defined.

\section{Evaluation Setting}
\label{sec:evaluation_setting}

In this section, we detail the settings for evaluating Word2API, including Research Questions (RQs), baseline algorithms, the evaluation strategy, and evaluation metrics.

\subsection{Research Questions}



\textbf{RQ1: How does Word2API perform against the baselines in relatedness estimation between a word and an API?}

To estimate term relatedness, many algorithms have been proposed.
We compare Word2API with these algorithms to show the effectiveness of Word2API.

\textbf{RQ2: How does Word2API perform under different settings?}

For generalization, Word2API utilizes the default settings of
the word embedding tool for vector generation.
This RQ evaluates Word2API under different parameter settings.

\textbf{RQ3: Does the shuffling step in training set creation contribute to the performance of Word2API?}

We investigate whether the shuffling strategy can better train word and API vectors.

\subsection{Baseline Algorithms for relatedness estimation}
This part explains the main algorithms for relatedness estimation \cite{mahmoud2015estimating}
and shows the baselines in this study.

\subsubsection{Latent Semantic Analysis (LSA)}
LSA (also called Latent Semantic Indexing) \cite{landauer1997solution}
first represents the documents in a corpus with an $m \times n$ matrix.
In the matrix, each row denotes a term in the corpus,
each column denotes a document,
and the value of a cell is the term weight in a document.
Then, LSA applies Singular Value Decomposition to transform and reduce the matrix into an $m \times n'$ matrix.
Each row of the matrix is an $n'$-dimensional vector that can be used to estimate the relatedness of different terms.

In this study, the inputs of LSA are word-API tuples.
We take each tuple as a document.
The value of a cell in the matrix is the frequency of a term in the document.
Due to the large number of tuples ($>$ 10 million),
we randomly sample 20\% tuples for training to resolve the computational problems in calculating high-dimensional matrices.
$n'$ is set to 200, since it achieves acceptable results on relatedness estimation \cite{mahmoud2015estimating}.
We implement LSA with Matlab.

\subsubsection{Co-occurrence based Methods}
Co-occurrence based methods assume that terms are semantically related
if they tend to co-occur in the same document or a fixed window size of the document.
In this experiment, a document means a word-API tuple.
Word2API belongs to this category.
Besides, we highlight several other representative algorithms,
including Pointwise Mutual Information (PMI), Normalized Software Distance (NSD), and Hyperspace Analogue to Language (HAL).

PMI measures term relatedness by comparing the probability of co-occurrence of two terms and the probability of occurrence of each term \cite{church1990word}.
Co-occurrence means two terms co-occur in the same document regardless of the position
and occurrence means a term occurs in a document.
PMI of a word $w$ and an API $a$ is defined as:
\begin{equation}
\begin{small}
\begin{aligned}
\label{pmi_formula}
{PMI(w, a)=\log\frac{p(w,a)}{p(w)p(a)}\approx \log\frac{f(w,a)}{(f(w)) \times (f(a))}},
\end{aligned}
\end{small}
\end{equation}
where $p(w,a)$ is the probability that $w$ and $a$ co-occur in a word-API tuple.
It can be estimated by $f(w,a)$, namely the number of tuples that contain both $w$ and $a$ divided by the total tuples' number.
$p(w)$ or $p(a)$ is the probability that $w$ or $a$ occurs in a tuple respectively,
which can be estimated by $f(w)$ or $f(a)$ similarly.

NSD \cite{mahmoud2015estimating} calculates the similarity between a word $w$ and an API $a$ with the following formula \cite{cilibrasi2007google}:
\begin{equation}
\begin{footnotesize}
\begin{aligned}
NSD(w,a)\!=\!\frac{\max \left \{ \log(f(w)),log(f(a)) \right \} \!-\! \log(\left | f(w)\bigcap f(a) \right |)}{\log(N)-\min\left \{ \log(f(w)),log(f(a)) \right \}},
\end{aligned}
\end{footnotesize}
\end{equation}
where $f(w)$ and $f(a)$ are the same definitions as those in formula (\ref{pmi_formula}) and $N$ is the number of tuples.

HAL \cite{lund1996producing} constructs a high dimensional $n \ast n$ matrix to represent the co-occurrences of all the $n$ terms in the word-API tuples.
Each cell (row$_i$, column$_j$) in the matrix is the weight between term$_{i}$ and term$_{j}$,
which is formalized as the Positive PMI (PPMI) between the corresponding terms \cite{tian2014automated}:
\begin{equation}
\begin{small}
\begin{aligned}
PPMI=
\left\{
            \begin{array}{lr}
             \!\!\!PMI({\textrm{\emph{term}}}_i,{\textrm{\emph{term}}}_j), & \!\!\!PMI({\textrm{\emph{term}}}_i,{\textrm{\emph{term}}}_j)>0 \\
             0 & otherwise\\
            \end{array}
\right.
\end{aligned}
\end{small}
\end{equation}

\subsubsection{Thesaurus-based Methods}
This line of methods uses linguistic dictionaries, e.g., WordNet,
for relatedness estimation.
However, such methods may be ineffective in software engineering areas \cite{mahmoud2015estimating,chen2017unsupervised},
due to the lack or mistaken definition of software-specific terms in the dictionaries, e.g., program reserved identifiers and APIs.
We do not take them as baselines.

\subsection{Evaluation Strategy}
As to our knowledge, no dataset is publicly available for word-API relatedness estimation,
as most evaluations depend on human judgements \cite{subramanian2014live,mahmoud2015estimating}.
We follow the widely accepted methodology of TREC\footnote{Text REtrieval Conference TREC. http://trec.nist.gov/}
for evaluation \cite{voorhees2001overview, yilmaz2015overview},
a popular Text REtrieval Conference of over 25 years' history.

Given a corpus,
TREC selects a set of queries for different algorithms to retrieve texts, e.g., web pages or documents.
The results are ranked in a descending order.
The top-k (usually, k=100 \cite{voorhees2001overview}) results are submitted to TREC.
TREC merges the results from different algorithms
and asks volunteers to judge the relatedness of the query-result pairs subjectively in a binary manner (related or unrelated).
Similar to TREC, we conduct the evaluation as follows.

\begin{table}[tp]
  \centering
  \fontsize{6.0}{6.8}\selectfont
  \begin{threeparttable}
  \caption{Selected words for evaluation}
  \label{tab:selected_words}
    \begin{tabular}{p{0.03cm}<{\centering}p{0.82cm}<{\centering}p{0.02cm}<{\centering}p{0.82cm}<{\centering}p{0.03cm}<{\centering}p{0.85cm}<{\centering}p{0.03cm}<{\centering}p{0.85cm}<{\centering}p{0.03cm}<{\centering}p{0.85cm}<{\centering}}
    \toprule
    \#&Word&\#&Word&\#&Word&\#&Word&\#&Word\cr
    \midrule
   1&agent    &11&delete     &21&key   &31&random &41&tail        \cr
   2&average  &12&display    &22&length&32&remote &42&thread      \cr
   3&begin    &13&environment&23&mp3   &33&request&43&timeout     \cr
   4&buffer   &14&file       &24&next  &34&reserve&44&transaction \cr
   5&capital  &15&filter     &25&node  &35&scale  &45&uuid        \cr
   6&check    &16&graphics   &26&object&36&select &46&validity    \cr
   7&classname&17&http       &27&open  &37&session&47&word        \cr
   8&client   &18&input      &28&parse &38&startup&48&xml         \cr
   9&current  &19&interrupt  &29&port  &39&string &49&xpath       \cr
   10&day     &20&iter       &30&post  &40&system &50&year        \cr
      \bottomrule
    \end{tabular}
    \end{threeparttable}
\end{table}

\subsubsection{Word selection}
This step selects a set of words as queries.
Initially, we randomly select 50 words from the GitHub corpus.
Among these words, nouns and verbs are selected as queries
as they are more descriptive \cite{howard2013automatically}.
The other words are removed.
The removed words are replaced by other randomly selected nouns or verbs
until the number of words reaches 50 (in Table \ref{tab:selected_words}).
This number is comparable to TREC \cite{yilmaz2015overview}
and other experiments in software engineering \cite{gu2016deep,mahmoud2015estimating}.

This experiment selects 50 words for evaluation.
It is a direct way to evaluate the semantic relatedness between words and APIs as suggested by previous studies \cite{gracia2008web,mahmoud2015estimating}.
Since human usually have some intuitive understandings to APIs,
the evaluation helps us understand whether the results returned by each algorithm are in accordance with the human intuition.

\subsubsection{API collection}
We run Word2API and the baseline algorithms with the selected words.
For each word, we collect and merge the top-100 recommended APIs for evaluation.

\subsubsection{Human judgement}
Theoretically, there are 25,000 word-API pairs for judgements (50 words$\times$5 algorithms$\times$100 recommendations).
Since some APIs may be recommended by more than one algorithm, there are 19,298 judgements eventually.
Due to the large number of word-API pairs,
we follow TREC to randomly split them into three non-overlapping partitions and assign the partitions to three volunteers
for evaluation (related or unrelated).
Each volunteer evaluates about 6,433 word-API pairs.
The volunteers are graduate students,
who have 3-5 years' experience in Java.
We take them as junior developers.
Since Junior developers (less than 5 years' experience) inhabit over 50\% of all developers according to a survey\footnote{https://insights.stackoverflow.com/survey/2016}
of 49,521 developers in Stack Overflow,
the evaluation may be representative to the view of many developers.

The definition of relatedness is open \cite{mahmoud2015estimating}.
Volunteers could consider the linguistic definition of a word, the usage scenarios of an API, etc.
We ask volunteers to record how they understand each word during evaluation,
i.e., the definition that they evaluate the word-API pairs.
The judgements take 2 weeks.
On average,
86 APIs are considered to be related to a word.

To evaluate the validity of human judgements,
we randomly select a statistically significant sample for re-evaluation based on the total number of 19,298 word-API pairs
with a confidence level of 99\% and a confidence interval of 5\% \cite{Maldonado2017Using},
resulting in a sample of 644 word-API pairs.
We send the sample to a new volunteer for judgements.
The Cohen's Kappa coefficient \cite{cohen1960coefficient} between the first and second round of judgements is 0.636,
which means that volunteers substantially agree on the judgements.

\subsection{Evaluation Metrics}
Based on the human judgements,
we evaluate each algorithm from two aspects,
namely, given a word,
how many related APIs can be correctly recommended
and whether the related APIs are ranked higher than the unrelated ones.
For these aspects, precision and NDCG are employed \cite{mcmillan2011portfolio,nie2016query}.
\begin{equation}
\begin{small}
\begin{aligned}
Precision@k=  \frac{\textrm{\# of relevant APIs to word}_i\textrm{ in top-}k}{k},
\end{aligned}
\end{small}
\end{equation}
\begin{equation}
\begin{footnotesize}
\begin{aligned}
NDCG@k=\frac{DCG@k}{IDCG@k}\quad
(DCG@k=\sum _{i=1}^{k}{\frac{r_i}{\log_2 i+1}}),
\end{aligned}
\end{footnotesize}
\end{equation}
where $r_i=1$ if the $i$th API is related to the given word, and $r_i=0$ otherwise.
IDCG is the ideal result of DCG, which all related APIs in a ranking list rank higher than the unrelated ones.
For example, if an algorithm recommends five APIs in which the 2nd, 4th APIs are related,
we can represent the results as \{0,1,0,1,0\}.
Then the ideal result is \{1,1,0,0,0\}.

\section{Evaluation Results}
\label{sec:evalution_results}
\subsection{Answer to RQ1: Baseline Comparison}

\subsubsection{Precision and NDCG}

\begin{figure}
\centering
  \subfigure[Evaluation on precision]{
  \label{fig:relatedness_p}
  \includegraphics[width=3.99cm]{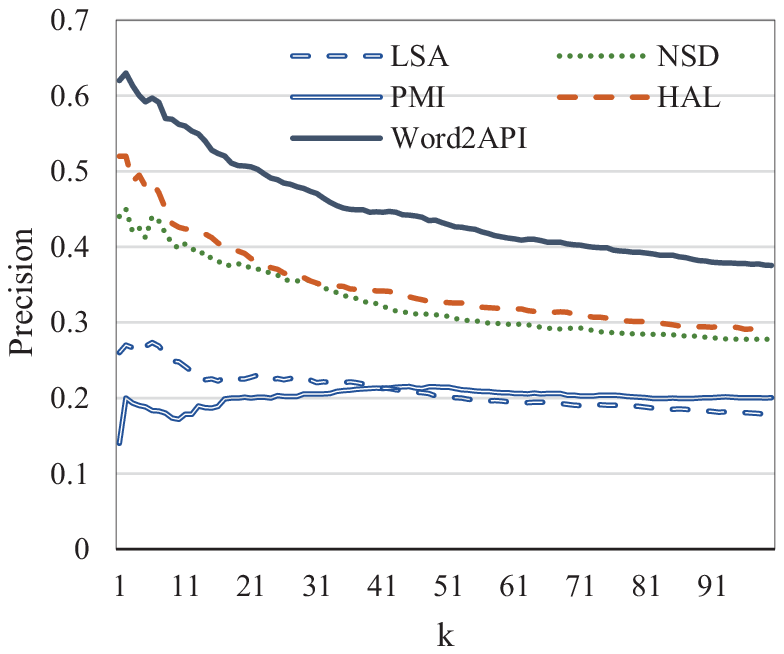}}
  \subfigure[Evaluation on NDCG]{
  \label{fig:relatedness_ndcg}
  \includegraphics[width=3.99cm]{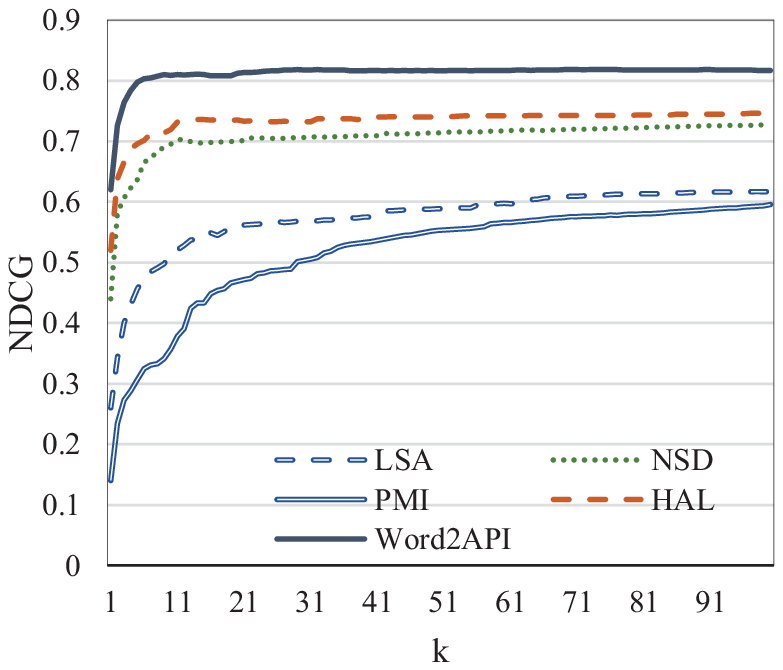}}
  \caption{Precision and NDCG on 50 selected words.}
  \label{fig:relatedness_p_ndcg}

\end{figure}

\begin{table*}[tp]
  \centering
  \fontsize{5.8}{8.5}\selectfont
  \begin{threeparttable}
  \caption{Examples of top-10 recommended APIs for different algorithms.}
  \label{tab:api_examples}
    \begin{tabular}{p{0.45cm}p{2.3cm}p{3.2cm}p{2.8cm}p{3.0cm}p{3.8cm}}
    \toprule
    &LSI&PMI&NSD&HAL&Word2API\cr
    \midrule
    {\multirow{10}{*}{capital}}
&Character\#getType            &NSMException$^1$\#fillInStackTrace  &\textbf{Character\#toUpperCase}    &\textbf{Character\#toTitleCase}   &\textbf{Character\#toUpperCase}  \cr
&StringBuilder\#insert         &\textbf{Character\#toTitleCase}     &\textbf{Character\#toTitleCase}    &Character\#isTitleCase            &\textbf{Character\#toTitleCase}  \cr
&Pattern\#normalizeSlice       &Character\#offsetByCodePoints       &\textbf{Character\#isUpperCase}    &String\#getValue                  &\textbf{Character\#isUpperCase}  \cr
&Pattern\#normalizeClazz   &\textbf{Character\#toUpperCase}     &Character\#isLetter                &\textbf{Character\#isUpperCase}   &\textbf{Character\#toLowerCase}  \cr
&\textbf{Character\#toUpperCase}&\textbf{Character\#isUpperCase}    &\textbf{Character\#isLowerCase}    &\textbf{Character\#isTitleCaseImpl}&\textbf{Character\#isLowerCase}  \cr
&StringBuilder\#reverse      &ToLongBiFunction$<$T$>$\#applyAsLong&\textbf{Character\#toLowerCase}    &\textbf{Character\#isLowerCase}    &Character\#isTitleCase  \cr
&StringBuilder\#appendCP$^2$ &LongStream\#sum                     &Character\#offsetByCodePoints &CharacterData\#toTitleCase     &IndexedPropertyDescriptor\#setReadMethod \cr
&StringBuilder\#setLength    &\textbf{Vector$<$T$>$\#subList}       &StringBuilder\#setCharAt      &IAVException$^3$\#printStackTrace       &StringBuilder\#setCharAt\cr
&StringBuilder\#setCharAt   &Character\#isLetter                   &NSMException\#fillInStackTrace&ITException$^4$\#fillInStackTrace       &PropertyDescriptor\#setName \cr
&\textbf{Character\#isAlphabetic}&\textbf{Character\#isLowerCase}  &String\#codePointAt           &\textbf{Character\#toUpperCase}         &\textbf{String\#toUpperCase}     \cr
    \hline
    {\multirow{10}{*}{uuid}}
&Objects\#requireNonNull	&	\textbf{UUID\#.new}	&	JAXBException\#fillInStackTrace	&	\textbf{UUID\#toString}	&	\textbf{UUID\#randomUUID}	\cr
&Charset\#newEncoder	&	\textbf{UUID\#toString}	&	\textbf{UUID\#padHex}	&	\textbf{UUID\#.new}	&	\textbf{UUID\#toString}	\cr
&CharSequence\#equals	&	\textbf{UUID\#randomUUID}	&	\textbf{UUID\#md5}	&	\textbf{UUID\#randomUUID}	&	\textbf{UUID\#fromString}	\cr
&AssertionError\#.new	&	\textbf{UUID\#version}	&	\textbf{UUID\#makeUuid}	&	\textbf{UUID\#nameUUIDFromBytes}	&	\textbf{UUID\#getLeastSignificantBits}	\cr
&\textbf{UUID\#toString}	&	\textbf{UUID\#getMostSignificantBits}	&	\textbf{UUID\#generateUUIDString}	&	ThreadLocalRandom\#nextBytes	&	\textbf{UUID\#getMostSignificantBits}	\cr
&Supplier$<$T$>$\#get	&	\textbf{UUID\#getLeastSignificantBits}	&	\textbf{UUID\#digits}	&	\textbf{Random\#nextLong}	&	\textbf{UUID\#digits}	\cr
&Scanner\#hasNextShort	&	\textbf{UUID\#fromString}	&	\textbf{UUID\#version}	&	\textbf{UUID\#equals}	&	\textbf{UUID\#nameUUIDFromBytes}	\cr
&CharSequence\#charAt	&	Long\#intValue	&	\textbf{UUID\#.new}	&	IIOMetadataNode\#setNodeValue	&	SOAPEnvelope\#createQName	\cr
&NullPointerException\#.new	&	\textbf{UUID\#nameUUIDFromBytes}	&	\textbf{UUID\#timestamp}	&	Base64\#getUrlEncoder	&	\textbf{UUID\#makeUuid}	\cr
&TAccessor$^5$\#isSupported	&	\textbf{UUID\#digits}	&	\textbf{UUID\#nameUUIDFromBytes}	&	TemporalAdjusters\#previousOrSame	&	\textbf{UUID\#equals}	\cr
     \bottomrule
     \end{tabular}
    \end{threeparttable}

    \begin{flushleft}
    \qquad Note: \quad {$^1$}NSMException: NoSuchMethodException \quad {$^2$}appendCP: appendCodePoint \quad {$^3$}IAVException: InvalidAttributeValueException \quad {$^4$}ITException: InvocationTargetException \\
    \qquad\qquad\quad\;\; {$^5$}TAccessor: TemporalAccessor
    \end{flushleft}
\end{table*}

Fig. \ref{fig:relatedness_p} and Fig. \ref{fig:relatedness_ndcg} are the averaged precision and NDCG for different algorithms over the selected 50 words respectively.
The x-axis is the ranking list size $k$ from 1 to 100
and y-axis is precision and NDCG on varied $k$.

In Fig. \ref{fig:relatedness_p}, Precision@1 of Word2API is 62\%,
which means that Word2API can find a semantically related API in the top-1 recommendation for 31 out of 50 query words.
This result outperforms the best baseline algorithm by 10\%.
When recommending 20 APIs by Word2API, half of the APIs are semantically related.
If we recommend 100 APIs, the precision of Word2API is still nearly 40\%.
Since there are about 86 related APIs for a selected word,
the result means that Word2API finds nearly half of the related APIs.
For NDCG, Word2API is superior to the other algorithms.
NDCG@1, NDCG@2 and NDCG@6 of Word2API are 0.620, 0.726 and 0.803  respectively,
which outperform the baselines by 0.102 to 0.496.
We explore the statistical significance of the results with the paired Wilcoxon signed rank test over the entire ranking list,
i.e., Precision@100 and NDCG@100.

\emph{$\rm{H}_0$: There is no significant difference between the performance of two algorithms over an evaluation metric.}

\emph{$\rm{H}_1$: There is significant difference between the performance of two algorithms over an evaluation metric.}

Since there are four baseline algorithms, the significance level is set to \begin{small}$0.05 / 4 \!\!=\! 1.25\times10^{-2}$\end{small} after Bonferroni correction \cite{weisstein2004bonferroni}.
The p-values on Precision@100 are \begin{small}$5.17\times10^{-9},\, 7.52\times10^{-9},\, 4.63\times10^{-7},\, 7.53\times10^{-5}$\end{small}
when comparing Word2API with LSA, PMI, NSD, HAL respectively.
H$_0$ is rejected.
Word2API significantly outperforms all the baseline algorithms.
We also achieve the same conclusion for NDCG@100.
The p-values are \begin{small}$1.77\times10^{-8},\, 2.99\times10^{-9},\, 8.76\times10^{-6},\, 5.20\times10^{-4}$\end{small} for LSA, PMI, NSD, and HAL respectively.

For the baseline algorithms, HAL and NSD are the best, followed by LSA and PMI.
Both HAL and NSD have been
applied on software engineering tasks in previous studies \cite{tian2014automated, mahmoud2015estimating}.
The two algorithms conduct relatedness estimation with high-dimensional vectors \cite{tian2014automated} or predefined functions \cite{mahmoud2015estimating}.
The drawback of HAL and NSD is that they cannot refine the relatedness of two terms with other terms in the same context.
In contrast, Word2API recovers a term based on the vectors of nearby terms.
The recovering step is to mine and refine
a term with the knowledge of
its context.
Hence, Word2API performs better over different metrics.

\subsubsection{Examples of recommended APIs}

Table \ref{tab:api_examples} presents examples of the recommended APIs for
words `capital' and `uuid'.\footnote{Other recommended APIs are at https://github.com/softw-lab/word2api}
We omit the API package names for brevity.
An API in bold is a related API by human judgements.
Volunteers think the word `capital' is semantically related with APIs that perform operations on the capital letters or first words.
It is a concept that may be related to different API packages, e.g., `String\#toUpperCase' or `Character\#toUpperCase'.
`uuid' is considered to be related to APIs in the java.util.UUID package and some APIs for random number generation.
It is a concept mainly related to a concrete package.

As shown in Table \ref{tab:api_examples}, the results of Word2API show similar understandings with volunteers.
It associates `capital' with APIs of `Character\#toUpperCase', `Character\#toLowerCase', and `String\#toUpperCase'.
Although some related APIs are also detected by HAL, NSD and PMI,
these algorithms still find some unrelated APIs in the top-5 results, e.g., `String\#getValue'.
For the word `uuid', many algorithms associate this concept with the UUID package.
Word2API is among the best of these algorithms.
In contrast, HAL fails to analyze this concept.
Only half of the top-10 APIs are related to `uuid'.
The reason may be that HAL represents terms with high-dimensional vectors.
The dimension equals to the vocabulary size.
The high-dimensional representation increases the computation complexity which makes HAL unprecise \cite{ye2016word},
e.g., introducing noises and being dominated by dimensions with large entry values.

\textbf{Conclusion}. Word2API outperforms the baseline algorithms in capturing the word-API semantic relatedness.

\subsection{Answer to RQ2: Parameter Influence}
\label{sec:rq2}

There are two main parameters for vector generation, namely the window size $w$ and the vector dimension $v$.\footnote{Besides, we analyze the influence of the shuffling times, the tuple length, and the evaluation metrics in Sec. S3 to S5 of the supplement.}
This RQ generates variants of Word2API to evaluate the parameter influence.
For the variants (in RQ2 and RQ3), additional human judgements are conducted on the new recommended APIs that have not been judged before.

\subsubsection{Window Size}
\begin{figure*}[htb]
\centering
  \includegraphics[width=16.0cm]{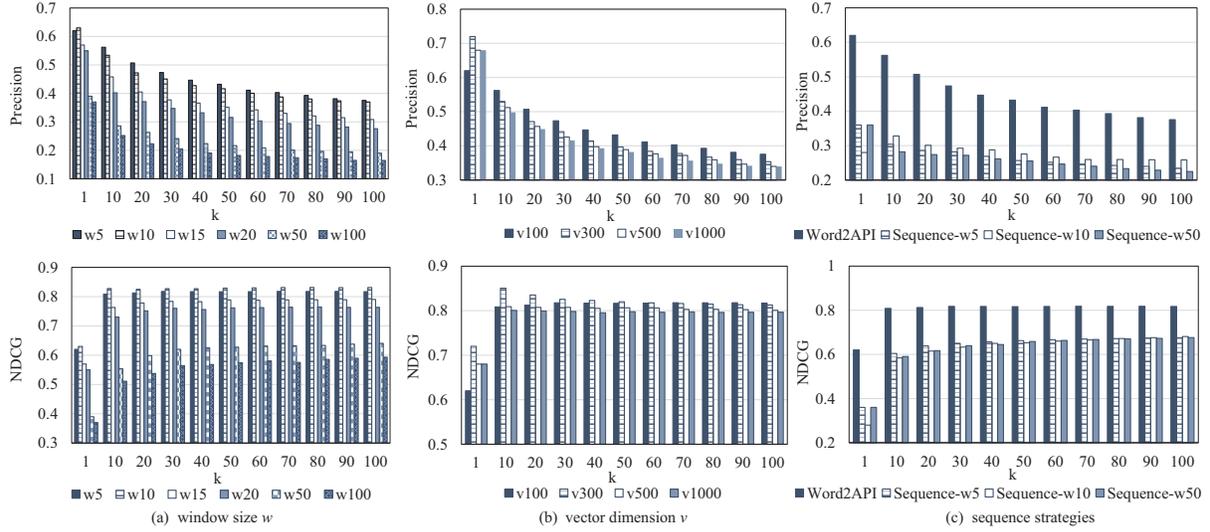}
  \caption{Precision and NDCG between Word2API and its variants.
  We tune the window size in (a).
  The vector dimension is evaluated in (b).
  We compare the shuffling strategy in Word2API with a sequence strategy in (c).}
  \label{fig:variants}
\end{figure*}

Fig. \ref{fig:variants}(a) shows precision and NDCG with respect to different window sizes.
We choose the window size varied from 5 to 100, including 5, 10, 15, 20, 50 and 100.
In the figures, the x-axis is the ranking list size $k$
and the y-axis is the corresponding precision or NDCG.
For simplicity, we only show the results of every ten ranking list size.

In Fig. \ref{fig:variants}(a), the precision of Word2API is stable when the window size is small.
The performance is nearly the same for $w=5$ and $w=10$.
For example, Precision@100 is 0.376 at $w=5$ and 0.370 at $w=10$.
If we increase $w$ to 50, the performance drops significantly.
The reason may be that,
Word2API constructs term vectors by maximizing the possibility to recover the current term vector with the co-occurred term vectors.
As the window size increases, the complexity of the recovering process increases,
which may become difficult for Word2API to recover the current term.

Similarly,
NDCG also tends to be stable when the window size is small.
We find that NDCG at $w=10$ is consistently better than that at $w=5$,
which means we can further improve Word2API by tuning the parameters.

\subsubsection{Vector Dimension}

We evaluate the influence of vector dimensions in Fig. \ref{fig:variants}(b).
The dimension is varied from 100 to 1000.
For the top-1 result,
the maximum margin of different dimensions is 0.100 on both precision and NDCG,
which happens between $v=100$ and $v=300$.
When Word2API recommends 100 APIs for a word,
the variation becomes small.
Precision@100 is 0.375 at $v=100$ and 0.340 at $v=1000$.
NDCG@100 is 0.817 at $v=100$ and 0.797 at $v=1000$.
We also average the differences between $v=100$ and $v=1000$ for the ranking list from 1 to 100.
The average difference between $v=100$ and $v=1000$ is 0.050 for precision and 0.018 for NDCG.
Hence, Word2API is relatively insensitive to the vector dimension overall.

The vector dimension determines the granularity to represent a term.
A small vector dimension means to represent a term with some abstract entries,
while a large vector dimension may generate more fine-grained vector representations.
Although a large vector dimension may better represent words and APIs,
it requires more data for training which slightly reduces Word2API's performance.
Hence, the overall ability of Word2API is not significantly affected.

\textbf{Conclusion}.  Word2API is stable at small window size
and relatively insensitive to the vector dimension.
We can improve Word2API by setting different parameters.

\subsection{Answer to RQ3: The Shuffling Strategy}
\label{sec:rq3}

\subsubsection{Comparison with the sequence strategy}
Word2API constructs word-API tuples from method comments and API calls to train word embedding.
It uses a shuffling strategy to obtain enough collocations between words and APIs in a word-API tuple.
In this subsection, we compare the shuffling strategy against a sequence strategy.
The sequence strategy combines the word sequence and the API sequence in a word-API tuple according to their original order,
i.e., words come before the APIs.
Then, it trains vectors on these combined data with the word embedding tool by the default parameters,
namely $w=5$, $v=100$, and -\emph{min}-\emph{count}$=5$.
We refer it as `Sequence-w5'.

We compare Word2API and Sequence-w5 in Fig. \ref{fig:variants}(c).
Sequence-w5 performs rather poor in estimating word-API relatedness.
Both Precision@1 and NCDG@1 are 0.360.
For top-100 recommended APIs, the precision and NDCG are 0.234 and 0.676 respectively,
In contrast, Word2API significantly outperforms Sequence-w5 by up to 26\% for both precision and NDCG.
The results demonstrate that the shuffling strategy
improves the ability of Word2API to construct vectors for semantically related words and APIs.

In addition, we increase the window size of Sequence-w5 to $w\!\!=\!\!10$ and $w\!\!=\!\!50$,
denoted as `Sequence-w10' and `Sequence-w50'.
The two variants investigate whether we can improve Sequence-w5 by increasing the window size.
As shown in Fig. \ref{fig:variants}(c), Sequence-w10 and Sequence-w50 perform similar to Sequence-w5.
For example,
Precision@100 are 0.2344 and 0.2248, and NDCG@100 are 0.6755 and 0.6761 for Sequence-w5 and Sequence-w50 respectively.
The differences are less than 0.01.
The reason may be that, although a large window size increases the number of co-occurred words and APIs for training word embedding,
it at the same time increases the computation complexity as discussed in Section \ref{sec:rq2}.
These two factors result in a stable performance of the sequence strategy.

\subsubsection{Comparison with the frequent itemset strategy}
This subsection compares the shuffling strategy with an alternative strategy,
namely the Frequent ItemSet (FIS) strategy, to generate a training set.
FIS takes each word-API tuple as a document and mines frequent itemsets with the Apriori algorithm.
To analyze the word-API relationship,
we collect the frequent 2-itemsets that contain a word and an API.
These word-API itemsets are considered to be highly related.
We calculate the confidence value from the word to the API in the frequent 2-itemsets.
After calculation, we traverse the 13,883,230 word-API tuples.
For an API in a word-API tuple, we search its highly related words in the same tuple
and put the API near the word with the largest word-to-API confidence value (on the right side of the word).
If the highly related words are not found, we leave the API at its original position.
We use these reordered word-API tuples to train word embedding.

There are two parameters for Apriori, i.e., the support value and the confidence value.
The support value is set to 0.0001.
We find each term in the word-API tuples appears in 1,491 tuples on average.
We consider an itemset to be frequent when all the terms in the itemset appear more frequently than the average value,
which attributes to a support value of 1,491/13,883,230, approximating to 0.0001.
At last, 48,961 frequent 2-itemsets are mined.
These itemsets contain 1,233 words.
Each word is related to 40 APIs on average.
We do not set a confidence value to further filter these itemsets,
because when the number of frequent itemsets is small, most word-API tuples are kept as their original order.

\begin{figure}
\centering
  \subfigure[Evaluation on precision]{
  \label{fig:itemset_p}
  \includegraphics[width=3.99cm]{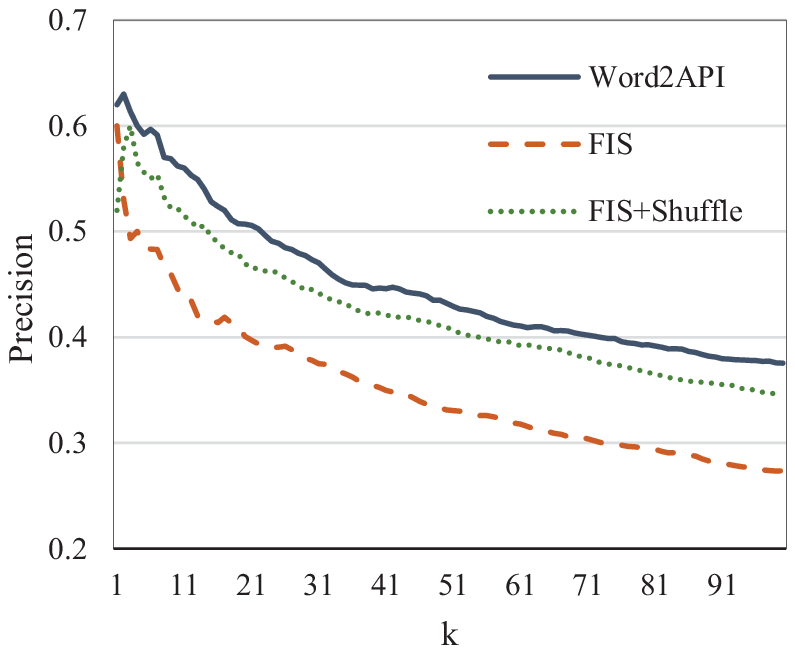}}
  \subfigure[Evaluation on NDCG]{
  \label{fig:itemset_ndcg}
  \includegraphics[width=3.99cm]{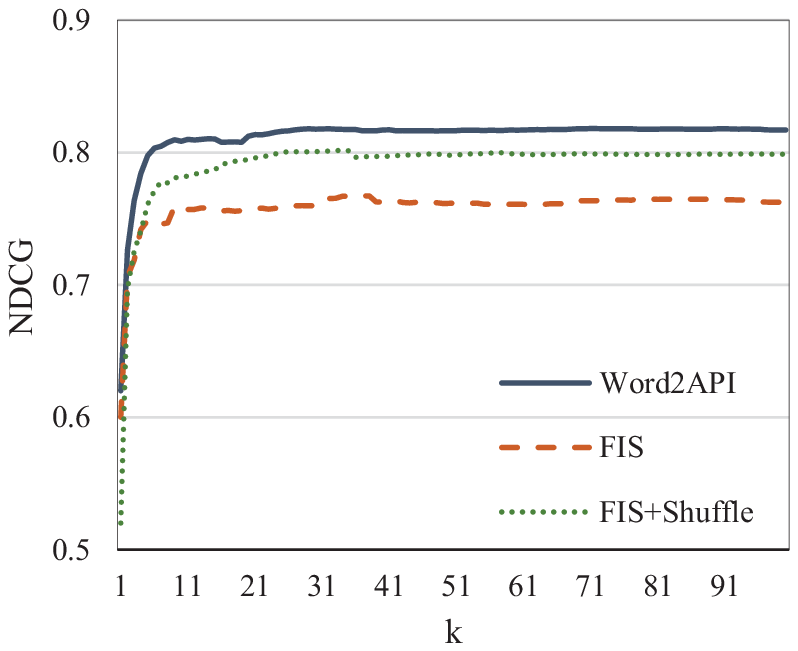}}
  \caption{Comparison on Shuffling and FIS strategies.}
  \label{fig:itemset_p_ndcg}

\end{figure}

As shown in Fig. \ref{fig:itemset_p_ndcg}, Word2API outperforms FIS by 0.02 to 0.102 in terms of precision and by 0.02 to 0.061 in terms of NDCG.
Although FIS is useful to generate the training set,
the shuffling strategy seems better than FIS.
The reason is that there are valuable information among all words and APIs in the same tuple.
When generating the training set with FIS,
the word embedding algorithm mainly mines the information among the highly related words and APIs instead of the overall information.

To prove this assumption, we propose another strategy named FIS+Shuffle.
This strategy first puts the highly related APIs near the word, and then shuffles the remaining words and APIs.
In Fig. \ref{fig:itemset_p_ndcg}, FIS+Shuffle improves FIS.
It means the shuffling strategy helps word embedding to analyze the overall information in a word-API tuple.
However, Word2API still outperforms FIS+Shuffle.
The reason may be that, for a word in frequent itemsets,
word embedding can hardly find the relationship between this word and every API,
as most surrounding APIs are limited to a few highly related ones.

\textbf{Conclusion}. The shuffling strategy improves the ability of Word2API to learn word-API relationships.

\section{Word2API for API Sequences Recommendation}
\label{sec:app1}

In Section \ref{sec:evalution_results}, we evaluate Word2API on relatedness estimation at the word-API level.
In the following parts, we further evaluate Word2API at the words-APIs level.
We show two typical applications of Word2API,
including API sequences recommendation and API documents linking.

\subsection{Overview}
The first application is API sequences recommendation.
It helps developers find APIs related to a short natural language query.
For example, if a developer searches for APIs implementing `generate random number',
APIs of `Random\#new, Random\#nextInt' may be recommended.


For a recommendation system, recent studies show that API based query expansion is effective to search related APIs \cite{lv2015codehow, raghothaman2016swim}.
Given a query, API based query expansion expands the query into an API vector.
Each entry of the vector is the probability or similarity that an API is related to the query.
The recommendation system uses the expanded API vector to search API sequences from a code base.


\subsection{Approach: API based Query Expansion}
We explain and compare the main algorithms for API based query expansion in this subsection,
including word alignment expansion (Align$_{Exp}$), API description expansion (Des$_{Exp}$) and Word2API expansion (Word2API$_{Exp}$).

\subsubsection{Word Alignment Expansion}
Align$_{Exp}$ \cite{raghothaman2016swim}
uses a statistical word alignment model \cite{brown1993mathematics} to calculate the probability between an API and a query.
The model is trained on alignment documents that consist of a set of words and related APIs \cite{raghothaman2016swim}.
We construct the alignment documents with word-API tuples \cite{gu2016deep}.
We use GIZA++\footnote{GIZA++. http://www.statmt.org/moses/giza/GIZA++.html} to implement
the word alignment model
and transform the query into a vector based on the probabilities.

\subsubsection{API Description Expansion}
Lv et al. expand a user query by analyzing the API descriptions \cite{lv2015codehow}.
Given a query, Des$_{Exp}$ collects all APIs and their descriptions in the Java SE API specification.
It calculates the similarity between the query and an API with a combined score of text similarity and name similarity.
Text similarity measures the similarity between the query and an API description by cosine similarity in the Term Frequency and Inverted Document Frequency (TF-IDF) space.
Name similarity splits an API into words by its camel style and measures the similarity between the query and the words.
Des$_{Exp}$ transforms the query into an API vector by the combined score.

\subsubsection{Word2API Expansion}
Given a user query (a set of words) and an API,
we apply the Words-APIs Similarity (formula \ref{fun:words_apis_similarity}) to calculate their similarity.
For each API, formula \ref{fun:words_apis_similarity} calculates the similarity between this API and each word in the query and then selects the largest value as the similarity between this API and the query.
In this way, we can get similarity values between the query and every API in the Java SE APIs.
Based on the similarities with all APIs,
we follow the previous study \cite{lv2015codehow} to select the top-10 APIs to expand the user query into an API vector.
The length of the vector is 10.
Each dimension of the vector represents an API.
The value of the dimension is the similarity between this API and the query.

After query expansion,
we employ a uniform framework to recommend API sequences \cite{raghothaman2016swim}.
This framework searches the word-API tuples to recommend APIs.
It transforms the APIs in each word-API tuple into a vector,
in which each entry means whether or not an API occurs in the current word-API tuple (0 or 1).
The length of this 0-1 vector is the number of Java SE APIs.
Then, it ranks the word-API tuples according to the cosine similarity between the expanded API vector and every 0-1 vector.
The framework finally returns the top-ranked word-API tuples.
Each tuple contains a set of APIs.
The order of these APIs is the same as that of in the word-API tuple.
The framework is efficient and naive to highlight the effect of different expansion approaches.

For this application,
the role of word2API is to calculate the similarity (relatedness) between a query and each API.
The similarities are used to expand a query into an API vector for searching word-API tuples.
We name this application as `API sequences recommendation', because each word-API tuple corresponds to an API sequence.
We find that word-API tuples not only have the APIs to implement a query,
but also introduce the context or examples on using the APIs,
as all word-API tuples are extracted from real-world source code.
For example, for the API `JFileChooser\#showOpenDialog' which implements `open file dialog',
the word-API tuples usually contain APIs of `JFileChooser\#new'  or `JFileChooser\#getSelectedFile'.
These APIs provide examples on what to do before or after using `JFileChooser\#showOpenDialog'.
Hence, comparing with other frameworks, e.g., deep neural network \cite{gu2016deep},
a retrieval based framework recommends valid and real-world API sequences,
that can be directly linked to diverse source code for understanding.
We compare Word2API with a deep neural network framework in Sec. S6 of the supplement.

\subsection{Evaluation: Query Expansion Algorithms}

\subsubsection{Motivation}
We compare Word2API$_{Exp}$ with Align$_{Exp}$ and Des$_{Exp}$ in recommending Java SE API sequences.

\subsubsection{Evaluation Method}
\label{sec:app1_method}
First, we evaluate these algorithms with 30 human written queries \cite{gu2016deep,raghothaman2016swim} listed in the first two columns of Table \ref{tab:app1_results}.
The evaluation is quantified with First Rank (FR) and Precision@k \cite{raghothaman2016swim}.
FR is the position of the first related API sequence to a query
and Precision@k is the ratio of related API sequences in the top-k results.
Similar to the previous study \cite{lv2015codehow},
two authors examined the results.
An API sequence is related if it contains the main API to implement a query
and receives related feedback from both authors.

Second, we conduct an automatic evaluation \cite{gu2016deep}
with 10,000 randomly selected tuples from all the word-API tuples.
We treat the word sequences in these tuples as queries
and the API sequences as the oracles.
The queries are used for an algorithm to search API sequences in the remaining tuples.
We compare sequence closeness between a recommended API sequence {\small $Seq_{rec}$} and the oracle sequence {\small $Seq_{orc}$} by BLEU score \cite{papineni2002bleu}:
\begin{equation}
\begin{footnotesize}
\begin{aligned}
BLEU\!\!&=\!\!BP\!\cdot\! \exp \left(\sum_{n=1}^{N}\!\!w_n\!\log\! \frac{{\textrm{\#n-grams in } Seq_{rec} \textrm{ and } Seq_{orc} \textrm{ +1}}}{{\textrm{\#n-grams in }} Seq_{rec} \textrm{+1}} \right )\\
BP&=\left\{\begin{matrix}
1 & \textrm{if } |Seq_{rec}|>|Seq_{orc}|\\
e^{1-|Seq_{orc}|/|Seq_{rec}|} & \textrm{if } |Seq_{rec}|\leq |Seq_{orc}|
\end{matrix}\right.
\end{aligned}
\end{footnotesize}
\end{equation}
where $|\cdot|$ is the length of a sequence, $N$ is the maximum gram number and $w_n$ is the weight of each type of gram.
According to previous studies \cite{gu2016deep, sutskever2014sequence}, $N$ is set to 4 and $w_n=1/N$.
It means that we calculate the overlaps of n-grams of {\small $Seq_{rec}$} and {\small $Seq_{orc}$} from 1 to 4 with equal weights.

For a ranking list of $k$ API sequences,
the BLEU score of the list is the maximum BLEU score between {\small $Seq_{rec}$} and {\small $Seq_{orc}$} \cite{gu2016deep}.
Since 10,000 tuples are used for evaluation,
we remove these tuples and their duplicate copies from the word-API tuples to re-train Word2API for fair comparison.

\subsubsection{Result}

\begin{table*}[pt]
     \setlength{\belowcaptionskip=1em}
  \centering
  \fontsize{5.5}{5.6}\selectfont
  \begin{threeparttable}
  \caption{Performance over 30 human written queries. P is short for Precision.}
  \label{tab:app1_results}
    \begin{tabular}{p{0.32cm}<{\centering}p{2.6cm}<{\centering}|p{0.35cm}<{\centering}p{0.35cm}<{\centering}p{0.35cm}<{\centering}|p{0.35cm}<{\centering}p{0.35cm}<{\centering}
    p{0.35cm}<{\centering}|p{0.35cm}<{\centering}p{0.35cm}<{\centering}p{0.35cm}<{\centering}|p{0.35cm}<{\centering}p{0.35cm}<{\centering}p{0.35cm}<{\centering}
    |p{0.35cm}<{\centering}p{0.35cm}<{\centering}p{0.35cm}<{\centering}|p{0.35cm}<{\centering}p{0.35cm}<{\centering}p{0.35cm}<{\centering}}
    \toprule
     \multirow{3}{*}{ID}& \multirow{3}{*}{Query}& \multicolumn{9}{c|}{Query Expansion}&\multicolumn{9}{c}{Search Engine}\cr
     &&\multicolumn{3}{c|}{Align$_{Exp}$\cite{raghothaman2016swim}}&\multicolumn{3}{c|}{Des$_{Exp}$ \cite{lv2015codehow}}&\multicolumn{3}{c|}{Word2API$_{\textrm{\emph{Exp}}}$}&\multicolumn{3}{c|}{Google$_{\textrm{\emph{GitHub}}}$}&\multicolumn{3}{c|}{Lucene$_{\textrm{\emph{API}}}$}&\multicolumn{3}{c}{Lucene$_{\textrm{\emph{API+Comment}}}$}\cr
   &&FR&P@5&P@10&FR&P@5&P@10&FR&P@5&P@10&FR&P@5&P@10&FR&P@5&P@10&FR&P@5&P@10\cr
    \midrule
Q1&	convert int to string         &	NF&	0&	0&	NF&	0&	0&	3&	0.2&	0.1&	6&	0&	0.1&	NF&	0&	0&	1&	0.4&	0.2\cr
Q2&	convert string to int         &	1&	1&	0.5&	NF&	0&	0&	1&	0.8&	0.8&	1&	0.8&	0.7&	NF&	0&	0&	8&	0&	0.3\cr
    Q3&	append string                 &	1&	1&	1&	1&	1&	1&	1&	1&	1&	7&	0&	0.2&	1&	1&	1&	1&	1&	1\cr
    Q4&	get current time              &	NF&	0&	0&	NF&	0&	0&	1&	1&	1&	1&	0.8&	0.5&	1&	1&	1&	1&	1&	0.8\cr
    Q5&	parse datetime from string    &	10&	0&	0.1&	NF&	0&	0&	1&	1&	0.7&	1&	1&	1&	NF&	0&	0&	NF&	0&	0\cr
    Q6&	test file exists              &	1&	1&	1&	1&	1&	1&	1&	0.8&	0.8&	2&	0.8&	0.9&	NF&	0&	0&	NF&	0.2&	0.1\cr
    Q7&	open a url                    &	1&	1&	1&	1&	1&	1&	1&	0.8&	0.8&	1&	0.2&	0.3&	1&	1&	1&	1&	1&	1\cr
    Q8&	open file dialog              &	NF&	0&	0&	1&	0.8&	0.7&	1&	0.4&	0.7&	1&	0.6&	0.5&	1&	1&	1&	3&	0.2&	0.3\cr
    Q9&	get files in folder           &	NF&	0&	0&	1&	0.8&	0.9&	1&	1&	0.9&	1&	0.6&	0.5&	NF&	0&	0&	NF&	0&	0\cr
    Q10&	match regular expressions    &	1&	1&	0.8&	1&	0.6&	0.7&	1&	1&	1&	2&	0.2&	0.5&	NF&	0&	0&	1&	1&	0.9\cr
    Q11&	generate md5 hash code       &	NF&	0&	0&	NF&	0&	0&	1&	1&	1&	1&	0.8&	0.6&	NF&	0&	0&	8&	0&	0.2\cr
    Q12&	generate random number       &	1&	0.4&	0.2&	1&	1&	1&	1&	1&	1&	1&	0.8&	0.6&	1&	0.6&	0.6&	1&	1&	0.9\cr
    Q13&	round a decimal value        &	NF&	0&	0&	2&	0.2&	0.1&	1&	0.8&	0.8&	10&	0&	0.1&	5&	0.2&	0.6&	1&	0.8&	0.8\cr
    Q14&	execute sql statement        &	NF&	0&	0&	NF&	0&	0&	2&	0.6&	0.5&	1&	1&	0.8&	1&	0.8&	0.9&	1&	0.8&	0.7\cr
    Q15&	connect to database          &	1&	1&	1&	NF&	0&	0&	1&	1&	1&	1&	0.8&	0.6&	NF&	0&	0&	NF&	0&	0\cr
 Q16&	create file    &	1&	1&	1&	1&	1&	1&	1&	1&	1&	1&	0.6&	0.7&	NF&	0&	0&	1&	0.4&	0.2\cr
 Q17&	copy file   &	1&	1&	1&	1&	1&	1&	1&	0.6&	0.5&	1&	0.8&	0.6&	NF&	0&	0&	4&	0.2&	0.1\cr
 Q18&	copy a file and save it to your destination path&	1&	1&	1&	2&	0.2&	0.3&	1&	0.8&	0.9&	7&	0&	0.2&	NF&	0&	0&	10&	0&	0.1\cr
 Q19&	delete files and folders in a directory&	1&	1&	1&	3&	0.6&	0.4&	4&	0.4&	0.4&	4&	0.4&	0.5&	NF&	0&	0&	1&	0.8&	0.4\cr
 Q20&	reverse a string&	NF&	0&	0&	NF&	0&	0&	NF&	0&	0&	1&	1&	1&	NF&	0&	0&	5&	0.2&	0.1\cr
 Q21&	create socket &	NF&	0&	0&	1&	0.6&	0.4&	1&	1&	0.9&	4&	0.2&	0.3&	1&	1&	0.7&	NF&	0&	0\cr
  Q22&	rename a file&	NF&	0&	0&	NF&	0&	0&	4&	0.4&	0.5&	1&	0.6&	0.3&	NF&	0&	0&	NF&	0&	0\cr
 Q23&	download file from url&	1&	1&	0.7&	1&	1&	1&	5&	0.2&	0.3&	1&	1&	0.7&	9&	0&	0.2&	5&	0.2&	0.5\cr
 Q24&	serialize an object&	1&	1&	1&	1&	1&	1&	1&	1&	1&	3&	0.2&	0.3&	4&	0.4&	0.2&	1&	0.6&	0.3\cr
 Q25&	read binary file      &	1&	1&	0.6&	1&	1&	1&	1&	0.8&	0.8&	2&	0.4&	0.4&	7&	0&	0.1&	2&	0.6&	0.7\cr
 Q26&	save an image to a file&	1&	1&	1&	1&	1&	1&	5&	0.2&	0.4&	1&	0.4&	0.4&	1&	1&	1&	4&	0.2&	0.6\cr
 Q27&	write an image to a file&	1&	1&	1&	1&	0.8&	0.6&	2&	0.4&	0.3&	1&	0.8&	0.8&	1&	1&	1&	1&	1&	1\cr
 Q28&	parse xml  &	NF&	0&	0&	NF&	0&	0&	1&	0.2&	0.3&	1&	0.6&	0.6&	5&	0.2&	0.1&	1&	0.2&	0.1\cr
 Q29&	play audio&	NF&	0&	0&	1&	0.8&	0.9&	1&	0.4&	0.5&	1&	0.6&	0.6&	6&	0&	0.2&	1&	0.2&	0.2\cr
 Q30&	play the audio clip at the specified absolute URL&	NF&	0&	0&	1&	1&	1&	1&	0.6&	0.4&	1&	1&	0.8&	4&	0.4&	0.7&	2&	0.8&	0.9\cr
\hline
\multicolumn{2}{l|}{Avg.}	&	5.633&	0.513&	0.463&	4.467&	0.547&	0.533&	\textbf{1.933}&	\textbf{0.68}&	\textbf{0.677}&	2.233&	0.567&	0.537&	6.767&	0.32&	0.343&	4.367&	0.427&	0.413\cr
\multicolumn{2}{l|}{\emph{p}}	&	0.002&	0.146&	0.04&	0.01&	0.135&	0.068&	*&	*&	*&	0.568&	0.109&	0.02&	\begin{tiny}\textless\end{tiny}0.001&	0.002&	0.003&	0.014&	0.009&	0.006\cr

    \bottomrule
    \end{tabular}

     \end{threeparttable}

\end{table*}

\begin{figure}
\centering
  \setlength{\belowcaptionskip}{-1.0em}
  \includegraphics[width=6.68cm,height=3.8cm]{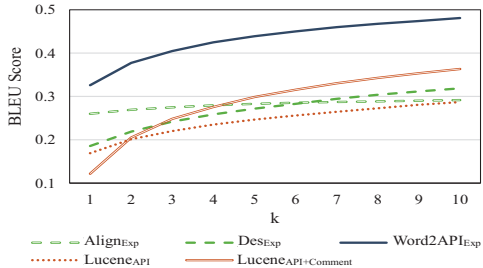}
  \caption{BLEU score for different expansion algorithms.}
  \label{fig:app1_bleu}
\end{figure}

\begin{table*}[pt]
     \setlength{\belowcaptionskip=1em}
     \centering
     \fontsize{5.5}{5.8}\selectfont
  \begin{threeparttable}
  \caption{Top-5 expanded APIs ranked by Word2API over 30 human written queries.}
  \label{tab:app1_examples}
    \begin{tabular}{p{2.95cm}p{2.9cm}p{3.2cm}p{3.25cm}p{3.38cm}}
    \toprule
    Ranked APIs&Ranked APIs&Ranked APIs&Ranked APIs&Ranked APIs\cr
    \midrule
Q1	&Q2	&Q3	&Q4	&Q5\cr
NFException$^1$\#toString	&NFException$^1$\#toString	&StringBuffer\#append	&System\#currentTimeMillis	&SimpleDateFormat\#parse\cr
Integer\#parseInt	&Integer\#parseInt	&StringBuilder\#append	&Date\#getTime	&DateFormat\#parse\cr
Object\#toString	&Object\#toString	&StringBuffer\#toString	&Clock\#millis	&GCalendar$^2$\#toZonedDateTime\cr
Byte\#parseByte	&Byte\#parseByte	&StringBuilder\#toString	&Calendar\#getTimeInMillis	&Date\#getField\cr
Integer\#byteValue	&Integer\#byteValue	&StringBuffer\#length	&BeanContext\#isDesignTime	&Calendar\#clear\cr
\hline
Q6	&Q7	&Q8	&Q9	&Q10\cr
File\#exists	&URL\#toString	&JFileChooser\#getSelectedFile	&File\#getName	&Pattern\#flags\cr
File\#isFile	&URL\#openStream	&JFileChooser\#showOpenDialog	&File\#getParentFile	&Pattern\#compile\cr
File\#canRead	&URL\#openConnection	&JFileChooser\#setDialogTitle	&File\#getPath	&Pattern\#pattern\cr
File\#isDirectory	&URLConnection\#getInputStream	&JFileChooser\#setCurrentDirectory	&File\#isFile	&Matcher\#toMatchResult\cr
File\#getPath	&URL\#toExternalForm	&JFChooser$^3$\#setFileSelectionMode	&File\#getAbsoluteFile	&Pattern\#matcher\cr
\hline
Q11	&Q12	&Q13	&Q14	&Q15\cr
MessageDigest\#digest	&SecureRandom\#nextInt	&Math\#round	&Statement\#close	&DataSource\#getConnection\cr
MessageDigest\#getInstance	&Random\#nextInt	&BigDecimal\#toPlainString	&Connection\#createStatement	&Connection\#close\cr
TreeMap\#hashCode	&Random\#nextDouble	&BigDecimal\#movePointRight	&SQLException\#getMessage	&Connection\#isClosed\cr
InvocationHandler\#hashCode	&Random\#nextBytes	&BigDecimal\#compactLong	&Connection\#close	&DriverManager\#getConnection\cr
NSAException$^4$\#printStackTrace	&SecureRandom\#nextBytes	&Math\#floor	&Connection\#prepareStatement	&Connection\#getNetworkTimeout\cr
\hline
Q16	&Q17	&Q18	&Q19	&Q20\cr
File\#exists	&File\#toPath	&File\#getParentFile	&File\#isDirectory	&StringBuilder\#reverse\cr
File\#toPath	&Files\#copy	&File\#mkdirs	&File\#exists	&Collections\#reverse\cr
File\#getAbsoluteFile	&File\#mkdirs	&File\#toPath	&File\#isFile	&StringBuffer\#reverse\cr
File\#getParentFile	&Arrays\#copyOf	&File\#getPath	&File\#getParentFile	&Collections\#sort\cr
Path\#toFile	&PFPermissions$^5$\#asFileAttribute	&Files\#copy	&File\#getPath	&Collections\#reverseOrder\cr
\hline
Q21	&Q22	&Q23	&Q24	&Q25\cr
SocketFactory\#getClass	&File\#renameTo	&URL\#toURI	&OutputStream\#hashCode	&DataInputStream\#close\cr
SSLSocket\#connect	&File\#delete	&URL\#toString	&OOStream$^6$\#dumpElementln	&BufferedInputStream\#read\cr
ServerSocket\#getChannel	&File\#getParentFile	&URL\#getFile	&Component\#doSwingSerialization	&File\#length\cr
SSLSocketFactory\#getDefault	&File\#exists	&URLConnection\#getContentLength	&Externalizable\#getClass	&FileInputStream\#close\cr
ServerSocketChannel\#setOption	&File\#getName	&URL\#getPath	&ObjectOutputStream\#writeObject	&File\#canRead\cr
\hline
Q26	&Q27	&Q28	&Q29	&Q30\cr
ImageIO\#write	&ImageIO\#write	&DocumentBuilder\#parse	&AudioSystem\#getLine	&AudioSystem\#getAudioInputStream\cr
File\#exists	&FileOutputStream\#close	&DBFactory$^7$\#newDocumentBuilder	&LUException$^8$\#printStackTrace	&Clip\#start\cr
FileOutputStream\#close	&File\#createTempFile	&DBFactory$^7$\#newInstance	&SourceDataLine\#start	&Clip\#open\cr
File\#getPath	&File\#mkdirs	&Document\#getDocumentElement	&Clip\#start	&AudioSystem\#getClip\cr
ImageIO\#read	&OutputStream\#close	&Element\#getAttribute	&AudioSystem\#getAudioInputStream	&Clip\#stop\cr
    \bottomrule
     \end{tabular}
    \end{threeparttable}

    \begin{flushleft}
     \quad {$^1$}NFException: NumberFormatException \quad {$^2$}GCalendar: GregorianCalendar \quad\; {$^3$}JFChooser: JFileChooser \quad\qquad\qquad\quad {$^4$}NSAException: NoSuchAlgorithmException   \\
     \quad {$^5$}PFPermissions: PosixFilePermissions \quad\;\;\, {$^6$}OOStream: ObjectOutputStream \quad {$^7$}DBFactory: DocumentBuilderFactory \quad {$^8$}LUException: LineUnavailableException \\
    \end{flushleft}

\end{table*}

Table \ref{tab:app1_results} shows the results for 30 human written queries.
`NF' means Not Found related APIs in the top 10 results.
We treat the FR value of `NF' as 11 to calculate the average FR \cite{gu2016deep}.
For 9 out of the 30 queries, all the query expansion algorithms can recommend related API sequences at top-1.
However, Align$_{Exp}$ fails to recommend APIs for many queries.
The reason is that, when Align$_{Exp}$ expands the query into an API vector,
the top ranked APIs in the vector are unrelated to the correct APIs.

The average FR of Word2API$_{Exp}$ is 1.933.
A related result is ranked at top-1 for 20 out of the 30 queries.
For precision, the average top-5 and top-10 precision by Word2API$_{Exp}$ is 0.680 and 0.677 respectively.
The results are superior to those of Align$_{Exp}$ and Des$_{Exp}$,
whose average top-10 precision values are 0.463 and 0.533 respectively.
Hence, by expanding queries with Word2API$_{Exp}$,
the recommendation framework achieves more related results on average than Align$_{Exp}$ and Des$_{Exp}$.
We conduct the paired Wilcoxon signed rank test on the 30 queries in the last row of Table \ref{tab:app1_results}.
When comparing Word2API against Align$_{Exp}$ and Des$_{Exp}$,
the p-values of FR, Precision@5 and Precision@10 are 0.0023, 0.1462, 0.0408, and 0.0117, 0.1347, 0.0675 respectively.

We report the results under 10,000 constructed queries in Fig. \ref{fig:app1_bleu}.
For the expansion approaches, Word2API$_{Exp}$ shows the best ability to transform queries into API vectors.
The BLEU scores are between 0.326 and 0.481,
which outperform Align$_{Exp}$ and Des$_{Exp}$ by up to 0.140 in terms of BLEU@1 and 0.189 in terms of BLEU@10.
The results pass the Wilcoxon test with p-value$<$0.025 after Bonferroni correction.
Since we take the same naive framework to retrieve API sequences,
it demonstrates that Word2API$_{Exp}$ makes the key contribution to the results.

Table \ref{tab:app1_examples} shows the top-5 expanded APIs by Word2API corresponding to each human written query.
In contrast to Table \ref{tab:api_examples}, this table reflects the ability of Word2API in understanding a set of words instead of a single one.
Among the top ranked APIs, most APIs are directly related to the queries.
For example, Word2API finds APIs of `Pattern\#compile, Pattern\#pattern, Pattern\#matcher' for query Q10 `match regular expressions'.
APIs of `Random\#nextInt, Random\#nextDouble, Random\#nextBytes' are ranked high for query Q12 `generate random number'.
With these APIs,
the naive framework can find more related API sequences on average
and rank the first related API sequence (FR) higher than the comparison algorithms.

Besides, we find that Word2API can interpret a query with APIs from multiple classes.
For example, in query Q4 `get current time', the top ranked APIs are `System\#currentTimeMillis', `Date\#getTime', and `Calendar\#getTimeInMillis'.
These APIs belong to different classes, including `java.lang.System', `java.util.Date', and `java.time.Clock'.
The same phenomenon can also be found in other queries, e.g., Q3 `append string', Q15 `connect to database', etc.
It means that Word2API may help developers understand a query by providing diverse APIs.

\subsubsection{Conclusion}
With the Word2API-expanded query, a system for API sequences recommendation significantly outperforms the comparison ones
in terms of FR and BLEU score.

\subsection{Evaluation: General-purpose Search Engines}
\label{sec:general_purpose}

\subsubsection{Motivation}
This section evaluates the performance of general-purpose search engines on recommending query-related APIs.

\subsubsection{Evaluation Method}
We propose three search engine based methods

{Google$_{\textrm{\emph{GitHub}}}$}.
This method searches a user query with Google search engine and collects the APIs in the top-10 web pages as results.
Since this study uses Java projects on GitHub to evaluate the ability of algorithms to match user queries with APIs,
for fair comparison, we limit Google$_{\textrm{\emph{GitHub}}}$ to search the resources on GitHub by rewriting a query as `java \emph{query} site:github.com'.

{Lucene$_{\textrm{\emph{API}}}$}.
The second method uses Lucene to search API sequences.
Lucene is a widely used open-source search engine
that uses text matching on words in queries and target documents to find related documents \cite{lucene}.
{Lucene$_{\textrm{\emph{API}}}$} takes the API sequences in word-API tuples as documents.
We first split each API in a word-API tuple into a set of words by its camel style.
Then, stop words removal and stemming are performed on the split words.
After that, we index these words as a document for search.

{Lucene$_{\textrm{\emph{API+Comment}}}$}.
The third method provides more knowledge to Lucene for accurate search.
Besides the words in API sequences,
{Lucene$_{\textrm{\emph{API+Comment}}}$} also indexes the words in the word sequence of a word-API tuple.
Since Word2API also uses API calls and method comments to mine word-API relationships,
this method helps us understand whether general-purpose search engines can better mine the semantic relatedness
when provided with the same amount of information.

We use the 30 human queries and 10,000 automatically constructed queries for evaluation.
We do not evaluate {Google$_{\textrm{\emph{GitHub}}}$} with the 10,000 queries
due to the network problem of automatically sending 10,000 queries to Google.
Since {Google$_{\textrm{\emph{GitHub}}}$} only returns web pages instead of Java APIs,
we use the following principle to evaluate {Google$_{\textrm{\emph{GitHub}}}$}.
We label a web page as correct, if the web page:
\begin{itemize}
  \item contains APIs related to the query, even though the APIs are from non-core Java APIs or other programming languages, e.g., Groovy and Scala;
  \item does not contain related APIs, but it implements a new method related to the query;
  \item does not contain source code, e.g., issue reports,
  but it contains API-like words related to the query.
\end{itemize}

\subsubsection{Result}
In Table \ref{tab:app1_results},
the average FR, P@5 and P@10 of Word2API$_{Exp}$ are superior to those of {Google$_{\textrm{\emph{GitHub}}}$}.
When considering the p-values,
Word2API performs similar to {Google$_{\textrm{\emph{GitHub}}}$} in terms of FR and P@5,
but significantly outperforms {Google$_{\textrm{\emph{GitHub}}}$} in terms of P@10 (p$<$0.05).
We find that even though {Google$_{\textrm{\emph{GitHub}}}$} is limited to search GitHub resources,
the correct APIs for some queries can still be easily obtained by matching a query with web page titles.
Hence, the results of {Google$_{\textrm{\emph{GitHub}}}$} may be attributed to both understanding the word-API relationships and leveraging the tremendous knowledge on the Internet.
Since Word2API$_{Exp}$ does not aim to recommend APIs with all the knowledge on the Internet,
we conclude that Word2API$_{Exp}$ achieves similar or better results compared to {Google$_{\textrm{\emph{GitHub}}}$} by only analyzing word-API relatedness.

For Lucene based methods,
Word2API$_{Exp}$ significantly outperforms {Lucene$_{\textrm{\emph{API}}}$} in terms of FR, P@5 and P@10.
It means the semantic gaps between words and APIs hinder the performance of general-purpose search engines in searching APIs by words.
When we provide more knowledge for Lucene, i.e., both API calls and method comments,
the performance of {Lucene$_{\textrm{\emph{API+Comment}}}$} improves.
However, Word2API$_{Exp}$ still outperforms {Lucene$_{\textrm{\emph{API+Comment}}}$}.
Since Word2API$_{Exp}$ and {Lucene$_{\textrm{\emph{API+Comment}}}$} use the same information to mine word-API relationships,
it means Word2API can better mine the semantic relatedness compared to a general-purpose search engine Lucene in this case,
when provided with the same amount of knowledge.

\subsubsection{Conclusion}
The semantic gaps hinder the performance of search engines in understanding APIs.
Word2API analyzes word-API relationships better than the search engine Lucene when provided with the same amount of knowledge.

\section{Word2API for API Documents Linking}
\label{sec:app2_linking}

\subsection{Overview}
The second application is API documents linking \cite{ye2016word} which analyzes the relationships between API documents and the questions in Q\&A (Question \& Answer) communities, e.g., Stack Overflow.
This application is more complex, since it needs to estimate semantic relatedness between a set of words and APIs, instead of a single API each time.

In Q\&A communities, participators discuss technical questions by replying and scoring to each other.
Given a newly submitted question, participators usually discuss and comprehend it with some APIs.
A statistic shows that more than one third (38\%) answers in Stack Overflow have at least one API \cite{parnin2012crowd}.
Hence, linking API documents to newly submitted questions may save participators' time to answer the questions \cite{ye2016word}.
In this part, we link questions in Stack Overflow to the documents in Java SE API specification \cite{ye2016word}.


\subsection{Approach: API Documents Linking}
Give a newly submitted question,
we introduce four typical algorithms to recommend related API documents.

\subsubsection{Vector Space Model (VSM)}
VSM transforms the question and API documents into vectors,
in which each entry is a word weighted by the TF-IDF strategy.
Then it ranks API documents by calculating the cosine similarity between the question vector and API document vectors.
We split the APIs and API-like words in these texts by the camel style to increase the number of matched words.

\subsubsection{Standard Word Embedding (WE)}
Ye et al. train word vectors for relatedness estimation with a standard word embedding model \cite{ye2016word}.
The vectors are generated by analyzing the words in Java and Eclipse API specifications, user/developer guides, and tutorials.
To link a question with API documents,
they transform the question and each API document into two words sets.
Then, they calculate the similarities of the word sets with the word vectors in a similar way as Formula \ref{fun:words_apis_similarity} (Words-APIs Similarity),
which replaces the word set and API set in Formula \ref{fun:words_apis_similarity} with two word sets.
For fair comparison, we also add the word sequences in word-API tuples for training.
WE is trained by the default parameters of the word embedding tool.

\subsubsection{Word2API Approach (Word2API)}
Word2API first extracts the words from the question and the method level APIs of an API type from each API document.
Then, it calculate the relatedness between the word set and API set by the Words-APIs Similarity.

\subsubsection{Integrated Approaches}
Previous studies also integrate VSM and WE to generate an integrate approach \cite{ye2016word}.
Given a question and an API document, we denote the similarity calculated by VSM, WE and Word2API as \begin{small}$Sim_{\rm{VSM}}$\end{small}, \begin{small}$Sim_{\rm{WE}}$\end{small}, and \begin{small}$Sim_{\rm{Word2API}}$\end{small} respectively.
We rank API documents by two types of integrations, namely VSM-WE (\begin{small}$Sim_{\rm{VSM\textrm{-}WE}}$\end{small})\cite{ye2016word} and VSM-Word2API (\begin{small}$Sim_{\rm{VSM\textrm{-}Word2API}}$\end{small}).
\begin{equation}
\begin{small}
\begin{aligned}
Sim_{{\rm{VSM\textrm{-}WE}}} \!=\! \alpha \!\!\times\!\! Sim_{{\rm{VSM}}} \!\!+\!\! (1\!\!-\!\!\alpha) \!\!\times\!\! Sim_{{\rm{WE}}},
\end{aligned}
\end{small}
\end{equation}
\begin{equation}
\begin{small}
\begin{aligned}
Sim_{{\rm{VSM\textrm{-}Word2API}}} \!=\! \alpha \!\!\times\!\! Sim_{{\rm{VSM}}} \!\!+\!\! (1\!\!-\!\!\alpha) \!\!\times\!\! Sim_{{\rm{Word2API}}},
\end{aligned}
\end{small}
\end{equation}
where $\alpha$ is the weight of different approaches.
The values are 0.18 and 0.36 for \begin{small}$Sim_{{\rm{VSM\textrm{-}WE}}}$\end{small} and \begin{small}$Sim_{{\rm{VSM\textrm{-}Word2API}}}$\end{small} respectively as we will discuss later.

\subsection{Evaluation}

\subsubsection{Motivation}
We evaluate the effectiveness of Word2API against the comparison algorithms on API documents linking.

\subsubsection{Evaluation Method}
We follow Ye et al. \cite{ye2016word} to construct a benchmark for evaluation.
We download Java tagged questions in Stack Overflow between August 2008 and March 2014,
since these questions have stabilized, i.e., no more edits are likely to be done.
Then we select a question,
if the score of the question exceeds 20, the score of its `best/accepted' answer exceeds 10,
and the `best/accepted' answer has at least one link to the Java SE API specification \cite{ye2016word}.
According to the criteria, 555 questions are collected.
We partition these questions into two parts.
The first 277 questions form a training set and the latter part is a testing set.
The size of the testing set is similar to the previous study \cite{ye2016word}.
We use the training set to tune the parameter $\alpha$ of the integrated approaches.
For an approach, we traverse $\alpha$ from 0.01 to 1.0 with a stepwise 0.01 and take the value that maximizes MAP in Equ. \ref{equ:map} as the final parameter value.
We take the API documents linked in the best/accepted answer as the oracle for evaluation.

For the testing set,
we compare the oracle API documents and the top-10 recommended API documents by MAP and MRR \cite{ye2016word}.
MAP is the mean of the average precision for each question.
\begin{equation}
\label{equ:map}
\begin{small}
\begin{aligned}
MAP &= \frac{1}{|Q|}\sum_{i=1}^{Q} AvgP_i \\
AvgP &= \sum_{k=1}^{N} r_k * Precision@k
\end{aligned}
\end{small}
\end{equation}
where $N$ is the number of recommended API documents for a question,
Precision@$k$ is the ratio of correctly recommended API documents in the top-k results,
and $r$ is a flag that $r_k$= 1 if the $k$th result is correct and $r_k$= 0 otherwise.

MRR is the mean reciprocal rank of the first correctly recommended API document for each question.
\begin{equation}
\begin{small}
\begin{aligned}
MRR = \frac{1}{|Q|}\sum_{i=1}^{Q} \frac{1}{FR_i},
\end{aligned}
\end{small}
\end{equation}
where $|Q|$ is the number of questions in the testing set and $FR_i$ is the position of the first related API document for $Q_i$.

\subsubsection{Result}
\begin{table}[tp]
  \centering
  \fontsize{7.0}{9.0}\selectfont
  \begin{threeparttable}
  \caption{MAP and MRR comparison.}
  \label{tab:app2_results}
    \begin{tabular}{p{2.7cm}p{1.2cm}p{1.2cm}}
    \toprule
    \quad Algorithms&MAP&MRR\cr
    \midrule
    \quad VSM	             &0.232	&0.259  \cr
    \quad WE \cite{ye2016word}	     &0.313	&0.354  \cr
    \quad Word2API	     &0.402	&0.433  \cr
    \hline
    \quad VSM+WE \cite{ye2016word}	 &0.340	&0.380  \cr
    \quad VSM+Word2API	 &0.436	&0.469  \cr
     \bottomrule
     \end{tabular}
    \end{threeparttable}
\end{table}

Table \ref{tab:app2_results} presents MAP and MRR of different algorithms.
Among the three atomic algorithms, the embedding based algorithms (WE and Word2API) are superior to VSM.
They improve VSM by up to 0.170 and 0.174 over MAP and MRR respectively.
The results show that sematic relatedness calculated by word embedding based algorithms are better than simple text matching (VSM) for this task.
For the two embedding based algorithms, Word2API performs better.
The results of MAP and MRR for Word2API are 0.402 and 0.433 over the testing set,
which outperform WE by 0.089 and 0.079 respectively.
It means that Word2API is more effective in mining semantic relatedness between words and APIs than WE,
which treats APIs as words.

We also find that text matching based algorithm VSM and embedding based algorithms can reinforce each other,
since they measure documents from different perspectives.
When we integrate the two types of algorithms, the results have an improvement by around 3\%,
e.g., VSM+Word2API reaches 0.436 on MAP and 0.469 on MRR.

Additionally, we note that some fine-grained text analysis techniques may further improve API documents linking,
e.g.,
deducing the APIs in source code snippets of the questions \cite{subramanian2014live,Treude2016Augmenting}.
We discuss this observation in Sec. S8 of the supplement.
The fine-grained analysis further improves API documents linking by nearly 5\%.

\subsubsection{Conclusion}
Word2API is superior to VSM and WE on relatedness estimation for API documents linking.

\section{Threats to Validity}
\label{sec:threats}

\noindent \textbf{Construction Validity}.
Word2API may require a large number of word-API tuples to construct a model.
As a machine learning algorithm,
Word2API is trained on the historical knowledge of word-API relationships.
When there are only a few word-API tuples containing an API,
Word2API may not well learn the relationship between words and this API.
A deep analysis is conducted in Sec. S7 of the supplement.
However, as the prevalence of open source,
we can easily download thousands of source code containing specific APIs from code repositories, e.g., GitHub, Google Code, etc.
As our preliminary statistic on GitHub, more than 583,779 and 388,300 projects contains at least one Android API and C\# API respectively.
These projects may facilitate the training of Word2API for such target APIs.

In addition, there are also threats in the two applications of Word2API.
We automatically select 10,000 word sequences to evaluate API sequences recommendation.
Since word sequences in the comments are not exactly the same with human queries,
we also evaluate Word2API with 30 human written queries.

\noindent \textbf{External Validity}.
The first threat comes from the human judgement processes.
To evaluate the semantic relatedness between query words and APIs,
several human judgements are conducted.
The selected query words may be vague for evaluation or unrealistic in real scenarios.
Meanwhile, the judgements are subjective and may bring biases.
We have observed some mislabeled APIs in this process.
To alleviate biases, we follow the TREC strategy for human judgements.
A re-evaluation shows a substantially agreement on the judgements with the Kappa score of 0.636.
In addition, we share the human judgement results at https://github.com/softw-lab/word2api for research.

The second threat is the generality of Word2API.
In this study, we evaluate Word2API at the word-API level with 50 query words
and at the words-APIs level with two applications.
More applications need to be investigated in the future.
For generality, we only use the default parameters to train Word2API.
Experiments show that Word2API works well without a fine-grained parameter optimization.

\section{Related Work}
\label{sec:related_work}

\begin{table}[tp]
  \centering
  \fontsize{6.2}{7.5}\selectfont
  \begin{threeparttable}
  \caption{Overview of the related work.}
  \label{tab:related_work}
    \begin{tabular}{p{0.2cm}p{1.1cm}p{2.18cm}p{3.15cm}}
    \toprule
    & Type &Paper & Knowledge \cr
    \midrule
    \multirow{11}{*}{\rotatebox{90}{Semantic Rel. Estimation}}
        &\multirow{3}{*}{Application}&Gu et al. \cite{gu2016deep}           &API sequences recommendation  \cr
        &&Ye et al. \cite{ye2016word}   &API documents linking \cr
        &&Corley et al. \cite{corley2015exploring}  &Feature location \cr
        \cline{2-4}
        &\multirow{3}{*}{Rule-based}&Beyer et al. \cite{beyer2015synonym}           &Heuristic rules  \cr
        &&Howard et al. \cite{howard2013automatically}   &Part-Of-Speech \cr
        &&Yang et al. \cite{yang2014swordnet}           &Term Morphology  \cr
        \cline{2-4}
        &\multirow{5}{*}{Corpus-based}&Mahmoud et al. \cite{mahmoud2015estimating}    &LSA, PMI, NSD (\emph{RQ1}) \cr
        &&Tian et al. \cite{tian2014automated,tian2014sewordsim}&HAL (\emph{RQ1}) \cr
        &&Chen et al. \cite{chen2017unsupervised}        &Word embedding (\emph{APP2})\cr
        &&Ye et al. \cite{ye2016word}                    &Word embedding (\emph{APP2}) \cr
        &&Nguyen et al. \cite{nguyen2017exploring}       &API embedding (\emph{Section S2 of the supplement})\cr
    \hline
    \multirow{7}{*}{\rotatebox{90}{Query Expansion}}
        &\multirow{5}{*}{Word-based}&Wang et al. \cite{wang2014active}              &Relevance feedback  \cr
        &&Hill et al. \cite{hill2014nl, roldan2013conquer}    &Co-occurred words  \cr
        &&Lu et al. \cite{lu2015query}                   &WordNet  \cr
        &&Nie et al. \cite{nie2016query}                 &Stack Overflow  \cr
        &&Campbell et al. \cite{campbell2017nlp2code}&Stack Overflow  \cr
        \cline{2-4}
        &\multirow{2}{*}{API-based}&Lv et al. \cite{lv2015codehow}                 &Similarity with API des. (\emph{APP1}) \cr
        &&Raghothaman et al. \cite{raghothaman2016swim}  &Word alignment (\emph{APP1})  \cr
        \bottomrule
     \end{tabular}
    \end{threeparttable}
\end{table}

We summarize the related work in Table \ref{tab:related_work},
including semantic relatedness estimation and query expansion.
For the highly related works, we also mark the \emph{RQ} or \emph{APP}lication that we compare these algorithms.

\subsection{Semantic Relatedness Estimation}
Semantic gaps between words and APIs negatively affect many software engineering tasks,
e.g., API sequences recommendation \cite{gu2016deep}, API documents linking \cite{ye2016word}, feature location \cite{corley2015exploring}, etc.
In this study, we propose Word2API to analyze the relatedness between words and APIs in a fine-grained, task-independent way.
Such analysis is useful for developers to understand the APIs and source code.

In the field of fine-grained relatedness estimation,
Beyer et al. \cite{beyer2015synonym} propose nine heuristic rules to suggest synonyms for Stack Overflow tags.
Howard et al. \cite{howard2013automatically} and Yang et al. \cite{yang2014swordnet} infer software-based semantically-similar words
by comparing the part-of-speech (verbs and nouns) and common words in API names and comments.
These techniques rely on specific rules without analyzing word relationships.

Hence corpus-based methods are proposed.
Mahmoud et al. \cite{mahmoud2015estimating} find that corpus-based methods outperform other methods on relatedness estimation.
Tian et al. \cite{tian2014automated, tian2014sewordsim} leverage Hyperspace Analogue to Language (HAL) to construct word vectors.
Chen et al. \cite{chen2017unsupervised} utilize word embedding to infer software-specific morphological forms, e.g., Visual C++ and VC++.
Similar vectors are also constructed on Java/Eclipse tutorials and user guides \cite{ye2016word}.
Besides,
Nguyen et al. propose API embedding to represent APIs of different languages \cite{nguyen2017exploring}.

Word2API is a corpus-based method.
It outperforms previous algorithms in word-API relatedness estimation.

\subsection{Query Expansion}
We take code search and API sequences recommendation as representative examples to enumerate the work in query expansion.
Code search aims to return code snippets for a user query \cite{keivanloo2014spotting}.
These snippets are usually more domain specific than API sequences \cite{gu2016deep}.
In this study, we classify query expansion into word-based expansion and API-based expansion.

Word-based expansion transforms a natural language query into more meaningful words.
Wang et al. \cite{wang2014active} leverage relevance feedback to expand queries with words in manually selected documents.
Hill et al. \cite{hill2014nl, roldan2013conquer} expand a query with frequently co-occurred words in code snippets.
Beside, external knowledge is also important for query expansion.
Lu et al. \cite{lu2015query} reformulate a user query with synonyms generated from WordNet.
Code snippets from Stack Overflow are also used for expanding queries \cite{nie2016query, campbell2017nlp2code}.
However, only a small part of Stack Overflow questions contains complete code snippets \cite{parnin2012crowd}.
In addition, word-based expansion aims at enhancing poor or simple queries.
Yet, the gaps between natural languages and APIs still exist.

Therefore, recent studies propose API-based expansion to transform a user query into related APIs.
Lv et al. \cite{lv2015codehow} expand a query by the text similarity and the name similarity between the query and API descriptions.
The effectiveness of this algorithms largely depends on the quality of API descriptions.
Hence, Raghothaman et al. \cite{raghothaman2016swim} utilize statistical word alignment models to expand queries into APIs.

Word2API belongs to API-based expansion.
A comparison with previous studies shows that Word2API is effective in expanding queries into API vectors.
In the future, we plan to conduct a comprehensive comparison and investigate the synergy of different types of expansion algorithms.

\section{Conclusion and Future Work}
\label{sec:conclusion}

In this study, we present our attempt towards constructing low-dimensional representations for both words and APIs.
Our algorithm Word2API leverages method comments and API calls to analyze semantic relatedness between words and APIs.
Experiments show that Word2API is effective in estimating semantically related APIs for a given word.
We present two applications of Word2API.
Word2API is a promising approach for expanding user queries into APIs and link API documents to Stack Overflow questions.
In the future, we plan to employ Word2API for other programming languages and applications,
and investigate different functions to measure similarity in addition to Words-APIs Similarity used in this paper.


%

\appendices


\ifCLASSOPTIONcompsoc
  \section*{Acknowledgment}
We thank the volunteers for their contributions to the exhausted human judgements processes.
\else
  \section*{Acknowledgment}
\fi


\ifCLASSOPTIONcaptionsoff
  \newpage
\fi


\begin{thebibliography}{10}
\providecommand{\url}[1]{#1}
\csname url@samestyle\endcsname
\providecommand{\newblock}{\relax}
\providecommand{\bibinfo}[2]{#2}
\providecommand{\BIBentrySTDinterwordspacing}{\spaceskip=0pt\relax}
\providecommand{\BIBentryALTinterwordstretchfactor}{4}
\providecommand{\BIBentryALTinterwordspacing}{\spaceskip=\fontdimen2\font plus
\BIBentryALTinterwordstretchfactor\fontdimen3\font minus
  \fontdimen4\font\relax}
\providecommand{\BIBforeignlanguage}[2]{{%
\expandafter\ifx\csname l@#1\endcsname\relax
\typeout{** WARNING: IEEEtran.bst: No hyphenation pattern has been}%
\typeout{** loaded for the language `#1'. Using the pattern for}%
\typeout{** the default language instead.}%
\else
\language=\csname l@#1\endcsname
\fi
#2}}
\providecommand{\BIBdecl}{\relax}
\BIBdecl

\bibitem{nie2016query}
L.~Nie, H.~Jiang, Z.~Ren, Z.~Sun, and X.~Li, ``Query expansion based on crowd
  knowledge for code search,'' \emph{IEEE Trans. on Services Computing},
  vol.~9, no.~5, pp. 771--783, 2016.

\bibitem{Robillard2011A}
M.~P. Robillard and R.~Deline, ``A field study of {API} learning obstacles,''
  \emph{Empirical Software Engineering}, vol.~16, no.~6, pp. 703--732, 2011.

\bibitem{raghothaman2016swim}
M.~Raghothaman, Y.~Wei, and Y.~Hamadi, ``{SWIM}: synthesizing what {I} mean -
  code search and idiomatic snippet synthesis,'' in \emph{Proc. of the 38th
  Int'l Conf. on Softw. Eng.}\hskip 1em plus 0.5em minus 0.4em\relax ACM, 2016,
  pp. 357--367.

\bibitem{ye2016word}
X.~Ye, H.~Shen, X.~Ma, R.~Bunescu, and C.~Liu, ``From word embeddings to
  document similarities for improved information retrieval in software
  engineering,'' in \emph{Proc. of the 38th Int'l Conf. on Softw. Eng.}\hskip
  1em plus 0.5em minus 0.4em\relax ACM, 2016, pp. 404--415.

\bibitem{Baeza2011Modern}
R.~A. Baeza-Yates and B.~A. Ribeiro-Neto, \emph{Modern Information Retrieval
  the Concepts and Technology Behind Search}.\hskip 1em plus 0.5em minus
  0.4em\relax ACM Press Books, 2011.

\bibitem{lv2015codehow}
F.~Lv, H.~Zhang, J.-g. Lou, S.~Wang, D.~Zhang, and J.~Zhao, ``Codehow:
  Effective code search based on {API} understanding and extended boolean
  model,'' in \emph{30th IEEE/ACM Int'l Conf. on Automated Softw. Eng.}\hskip
  1em plus 0.5em minus 0.4em\relax IEEE, 2015, pp. 260--270.

\bibitem{gu2016deep}
X.~Gu, H.~Zhang, D.~Zhang, and S.~Kim, ``Deep {API} learning,'' in \emph{Proc.
  of the 2016 24th ACM SIGSOFT Int'l Symposium on Foundations of Softw.
  Eng.}\hskip 1em plus 0.5em minus 0.4em\relax ACM, 2016, pp. 631--642.

\bibitem{piccioni2013empirical}
M.~Piccioni, C.~A. Furia, and B.~Meyer, ``An empirical study of {API}
  usability,'' in \emph{ACM / IEEE Int'l Symposium on Empirical Softw. Eng. and
  Measurement}, 2013, pp. 5--14.

\bibitem{zhou2017analyzing}
Y.~Zhou, R.~Gu, T.~Chen, Z.~Huang, S.~Panichella, and H.~Gall, ``Analyzing
  {API}s documentation and code to detect directive defects,'' in \emph{Proc.
  of the 39th Int'l Conf. on Softw. Eng.}, 2017, pp. 27--37.

\bibitem{mahmoud2015estimating}
A.~Mahmoud and G.~Bradshaw, ``Estimating semantic relatedness in source code,''
  \emph{ACM Trans. on Softw. Eng. and Methodology}, vol.~25, no.~1, p.~10,
  2015.

\bibitem{landauer1997solution}
T.~K. Landauer and S.~T. Dumais, ``A solution to plato's problem: The latent
  semantic analysis theory of acquisition, induction, and representation of
  knowledge.'' \emph{Psychological Review}, vol. 104, no.~2, p. 211, 1997.

\bibitem{miller1995wordnet}
G.~A. Miller, ``{W}ord{N}et: a lexical database for {E}nglish,''
  \emph{Communications of the ACM}, vol.~38, no.~11, pp. 39--41, 1995.

\bibitem{nguyen2017exploring}
T.~D. Nguyen, A.~T. Nguyen, H.~D. Phan, and T.~N. Nguyen, ``Exploring {API}
  embedding for {API} usages and applications,'' in \emph{Proc. of the 39th
  Int'l. Conf. on Softw. Eng.}\hskip 1em plus 0.5em minus 0.4em\relax IEEE
  Press, 2017, pp. 438--449.

\bibitem{apidefinition}
Wikipedia, ``Application programming interface,''
  \url{https://en.wikipedia.org/wiki/Application_programming_interface}, 2017.

\bibitem{bloch2006how}
J.~Bloch, ``How to design a good {API} and why it matters,'' in \emph{Companion
  to the ACM Sigplan Symposium on Object-Oriented Programming Systems,
  Languages, and Applications}, 2006, pp. 506--507.

\bibitem{mikolov2013distributed}
T.~Mikolov, I.~Sutskever, K.~Chen, G.~Corrado, and J.~Dean, ``Distributed
  representations of words and phrases and their compositionality,''
  \emph{Advances in Neural Information Processing Systems}, vol.~26, pp.
  3111--3119, 2013.

\bibitem{mikolov2013efficient}
T.~Mikolov, K.~Chen, G.~Corrado, and J.~Dean, ``Efficient estimation of word
  representations in vector space,'' \emph{CoRR, abs/1301.3781}, 2013.

\bibitem{Kalliamvakou2016An}
E.~Kalliamvakou, G.~Gousios, K.~Blincoe, L.~Singer, D.~M. German, and
  D.~Damian, ``An in-depth study of the promises and perils of mining github,''
  \emph{Empirical Software Engineering}, vol.~21, no.~5, pp. 2035--2071, 2016.

\bibitem{javaoverriding}
Oracle, ``{Overriding and hiding methods},''
  \url{http://docs.oracle.com/javase/tutorial/java/IandI/override.html}, 2017.

\bibitem{pascarella2017classifying}
L.~Pascarella and A.~Bacchelli, ``Classifying code comments in {Java}
  open-source software systems,'' in \emph{Proc. of the 14th Int'l Conf. on
  Mining Software Repositories}.\hskip 1em plus 0.5em minus 0.4em\relax IEEE
  Press, 2017, pp. 227--237.

\bibitem{margaret30todo}
S.~Margaret-Anne, J.~Ryall, R.~I. Bull, D.~Myers, and J.~Singer, ``To{D}o or to
  bug: Exploring how task annotations play a role in the work practices of
  software developers,'' in \emph{Proc. of the 30th Int'l. Conf. on Softw.
  Eng.}, pp. 251--260.

\bibitem{howard2013automatically}
M.~J. Howard, S.~Gupta, L.~Pollock, and K.~Vijay-Shanker, ``Automatically
  mining software-based, semantically-similar words from comment-code
  mappings,'' in \emph{Proc. of the 10th Working Conf. on Mining Software
  Repositories}.\hskip 1em plus 0.5em minus 0.4em\relax IEEE Press, 2013, pp.
  377--386.

\bibitem{lotufo2015modelling}
R.~Lotufo, Z.~Malik, and K.~Czarnecki, ``Modelling the `hurried' bug report
  reading process to summarize bug reports,'' \emph{Empirical Software
  Engineering}, vol.~20, no.~2, pp. 516--548, 2015.

\bibitem{porter1980algorithm}
M.~F. Porter, ``An algorithm for suffix stripping,'' \emph{Program}, vol.~14,
  no.~3, pp. 130--137, 1980.

\bibitem{allamanis2015suggesting}
M.~Allamanis, E.~T. Barr, C.~Bird, and C.~Sutton, ``Suggesting accurate method
  and class names,'' in \emph{Proc. of the 2015 10th Joint Meeting on
  Foundations of Softw. Eng.}\hskip 1em plus 0.5em minus 0.4em\relax ACM, 2015,
  pp. 38--49.

\bibitem{tantithamthavorn2016automated}
C.~Tantithamthavorn, S.~McIntosh, A.~E. Hassan, and K.~Matsumoto, ``Automated
  parameter optimization of classification techniques for defect prediction
  models,'' in \emph{Proc. of the 38th Int'l Conf. on Softw. Eng.}\hskip 1em
  plus 0.5em minus 0.4em\relax IEEE, 2016, pp. 321--332.

\bibitem{mihalcea2006corpus}
R.~Mihalcea, C.~Corley, and C.~Strapparava, ``Corpus-based and knowledge-based
  measures of text semantic similarity,'' in \emph{National Conf. on Artificial
  Intelligence and the 18th Innovative App. of Artificial Intelligence Conf.},
  2006, pp. 775--780.

\bibitem{church1990word}
K.~W. Church and P.~Hanks, ``Word association norms, mutual information, and
  lexicography,'' \emph{Computational Linguistics}, vol.~16, no.~1, pp. 22--29,
  1990.

\bibitem{cilibrasi2007google}
R.~L. Cilibrasi and P.~M. Vitanyi, ``The {Google} similarity distance,''
  \emph{IEEE Trans. on Knowledge and Data Eng.}, vol.~19, no.~3, 2007.

\bibitem{lund1996producing}
K.~Lund and C.~Burgess, ``Producing high-dimensional semantic spaces from
  lexical co-occurrence,'' \emph{Behavior Research Methods, Instruments, \&
  Computers}, vol.~28, no.~2, pp. 203--208, 1996.

\bibitem{tian2014automated}
Y.~Tian, D.~Lo, and J.~Lawall, ``Automated construction of a software-specific
  word similarity database,'' in \emph{Conf. on the 2014 IEEE Softw.
  Maintenance, Reengineering and Reverse Eng.}, 2014, pp. 44--53.

\bibitem{chen2017unsupervised}
C.~Chen, Z.~Xing, and X.~Wang, ``Unsupervised software-specific morphological
  forms inference from informal discussions,'' in \emph{Proc. of the 39th Int'l
  Conf. on Softw. Eng.}\hskip 1em plus 0.5em minus 0.4em\relax IEEE Press,
  2017, pp. 450--461.

\bibitem{subramanian2014live}
S.~Subramanian, L.~Inozemtseva, and R.~Holmes, ``Live {API} documentation,'' in
  \emph{Proc. of the 36th Int'l. Conf. on Softw. Eng.}\hskip 1em plus 0.5em
  minus 0.4em\relax ACM, 2014, pp. 643--652.

\bibitem{voorhees2001overview}
E.~M. Voorhees and D.~Harman, ``Overview of {TREC} 2001.'' in \emph{TREC},
  2001.

\bibitem{yilmaz2015overview}
E.~Yilmaz, M.~Verma, R.~Mehrotra, E.~Kanoulas, B.~Carterette, and N.~Craswell,
  ``Overview of {TREC} 2015 tasks track.'' in \emph{TREC}, 2015.

\bibitem{gracia2008web}
J.~Gracia and E.~Mena, ``Web-based measure of semantic relatedness,'' in
  \emph{Int'l Conf. on Web Information Systems Engineering}.\hskip 1em plus
  0.5em minus 0.4em\relax Springer, 2008, pp. 136--150.

\bibitem{Maldonado2017Using}
E.~D.~S. Maldonado, E.~Shihab, and N.~Tsantalis, ``Using natural language
  processing to automatically detect self-admitted technical debt,'' \emph{IEEE
  Trans. on Softw. Eng.}, vol.~PP, no.~99, pp. 1--1, 2017.

\bibitem{cohen1960coefficient}
J.~Cohen, ``A coefficient of agreement for nominal scales,'' \emph{Educational
  and Psychological Measurement}, vol.~20, no.~1, pp. 37--46, 1960.

\bibitem{mcmillan2011portfolio}
C.~McMillan, M.~Grechanik, D.~Poshyvanyk, Q.~Xie, and C.~Fu, ``Portfolio:
  finding relevant functions and their usage,'' in \emph{Proc. of the 33rd
  Int'l. Conf. on Softw. Eng.}\hskip 1em plus 0.5em minus 0.4em\relax ACM,
  2011, pp. 111--120.

\bibitem{weisstein2004bonferroni}
E.~W. Weisstein, ``Bonferroni correction,'' \emph{Wolfram Research, Inc.},
  2004.

\bibitem{brown1993mathematics}
P.~F. Brown, V.~J.~D. Pietra, S.~A.~D. Pietra, and R.~L. Mercer, ``The
  mathematics of statistical machine translation: Parameter estimation,''
  \emph{Computational Linguistics}, vol.~19, no.~2, pp. 263--311, 1993.

\bibitem{papineni2002bleu}
K.~Papineni, S.~Roukos, T.~Ward, and W.-J. Zhu, ``{BLEU}: a method for
  automatic evaluation of machine translation,'' in \emph{Proc. of the 40th
  Annual Meeting on Association for Computational Linguistics}.\hskip 1em plus
  0.5em minus 0.4em\relax Association for Computational Linguistics, 2002, pp.
  311--318.

\bibitem{sutskever2014sequence}
I.~Sutskever, O.~Vinyals, and Q.~V. Le, ``Sequence to sequence learning with
  neural networks,'' vol.~4, pp. 3104--3112, 2014.

\bibitem{lucene}
Apache, ``Apache lucene,'' \url{http://lucene.apache.org/}, 2018.

\bibitem{parnin2012crowd}
C.~Parnin, C.~Treude, L.~Grammel, and M.~A. Storey, ``Crowd documentation:
  Exploring the coverage and the dynamics of {API} discussions on {S}tack
  {O}verflow,'' \emph{Georgia Institute of Technology, Tech. Rep}, 2012.

\bibitem{Treude2016Augmenting}
C.~Treude and M.~P. Robillard, ``Augmenting {API} documentation with insights
  from {S}tack {O}verflow,'' pp. 392--403, 2016.

\bibitem{corley2015exploring}
C.~S. Corley, K.~Damevski, and N.~A. Kraft, ``Exploring the use of deep
  learning for feature location,'' in \emph{2015 IEEE Int'l Conf. on Softw.
  Maintenance and Evolution}.\hskip 1em plus 0.5em minus 0.4em\relax IEEE,
  2015, pp. 556--560.

\bibitem{beyer2015synonym}
S.~Beyer and M.~Pinzger, ``Synonym suggestion for tags on {S}tack {O}verflow,''
  in \emph{IEEE Int'l Conf. on Program Comprehension}, 2015, pp. 94--103.

\bibitem{yang2014swordnet}
J.~Yang and L.~Tan, ``{SW}ord{N}et: Inferring semantically related words from
  software context,'' \emph{Empirical Software Engineering}, vol.~19, no.~6,
  pp. 1856--1886, 2014.

\bibitem{tian2014sewordsim}
Y.~Tian, D.~Lo, and J.~Lawall, ``{SEW}ord{S}im: software-specific word
  similarity database,'' in \emph{Companion Proc. of the 36th Int'l Conf. on
  Softw. Eng.}, 2014.

\bibitem{wang2014active}
S.~Wang, D.~Lo, and L.~Jiang, ``Active code search: incorporating user feedback
  to improve code search relevance,'' in \emph{ACM/IEEE Int'l Conf. on
  Automated Softw. Eng.}, 2014, pp. 677--682.

\bibitem{hill2014nl}
E.~Hill, M.~Roldanvega, J.~A. Fails, and G.~Mallet, ``{NL}-based query
  refinement and contextualized code search results: A user study,'' in
  \emph{Softw. Maintenance, Reengineering and Reverse Eng.}, 2014, pp. 34--43.

\bibitem{roldan2013conquer}
M.~Roldan-Vega, G.~Mallet, E.~Hill, and J.~A. Fails, ``{CONQUER}: A tool for
  {NL}-based query refinement and contextualizing code search results,'' in
  \emph{IEEE Int'l Conf. on Softw. Maintenance}, 2013, pp. 512--515.

\bibitem{lu2015query}
M.~Lu, X.~Sun, S.~Wang, D.~Lo, and Y.~Duan, ``Query expansion via {W}ord{N}et
  for effective code search,'' in \emph{IEEE Int'l Conf. on Software Analysis,
  Evolution and Reengineering}, 2015, pp. 545--549.

\bibitem{campbell2017nlp2code}
B.~A. Campbell and C.~Treude, ``{NLP2C}ode: Code snippet content assist via
  natural language tasks,'' in \emph{Tool Demo of 2017 IEEE Int'l Conf. on
  Softw. Maintenance and Evolution}, 2017.

\bibitem{keivanloo2014spotting}
I.~Keivanloo, J.~Rilling, and Y.~Zou, ``Spotting working code examples,'' in
  \emph{Proc. of the 36th Int'l Conf. on Softw. Eng.}\hskip 1em plus 0.5em
  minus 0.4em\relax ACM, 2014, pp. 664--675.

\end{thebibliography}
\end{document}